\documentclass[%draft,
				a4paper,fleqn,usenatbib,useAMS]{mnras}

\usepackage{newtxtext,newtxmath}
\usepackage[T1]{fontenc}
\usepackage{ae,aecompl}

\usepackage{graphicx}	% Including figure files
\usepackage{amsmath}	% Advanced maths commands
\usepackage{amssymb}	% Extra maths symbols
\usepackage{amsfonts}
\usepackage{tikz}
\usepackage[version=3]{mhchem}
\usepackage{datetime}
\usepackage{booktabs}
\usepackage{float}
\usepackage{gensymb}
\usepackage{wasysym}
\usepackage{amsfonts}
\usepackage{courier}
\usepackage{caption}
\usepackage{subcaption}
\usepackage{multicol}   % Multi-column entries in tables
\usepackage{multirow}
\usepackage{pdflscape}	% Landscape pages
\usepackage[para]{threeparttable}
\usepackage{pifont}
\newcommand{\cmark}{\ding{51}}%
\newdateformat{mydate}{\twodigit{\THEDAY}{ }\monthname[\THEMONTH], \THEYEAR}
\mydate
\newdate{reportdate}{06}{6}{2017}
\newdate{datadate}{02}{2}{2017}

\newcommand{\rom}[1]{%
  \textup{\uppercase\expandafter{\romannumeral#1}}%
}

\title[System stability and HZ companions in the TESS era]{Predicting multiple planet stability and habitable zone companions in the TESS era}
       
\author[M. T. Agnew]{
Matthew T. Agnew,$^{1}$ %\thanks{E-mail: mn@ras.org.uk (KTS)}
Sarah T. Maddison,$^{1}$ 
Jonathan Horner$^{2}$ and 
Stephen R. Kane$^{3}$
\\
$^{1}$Centre for Astrophysics and Supercomputing, Swinburne University of Technology, Hawthorn, Victoria 3122, Australia\\
$^{2}$Centre for Astrophysics, University of Southern Queensland, Toowoomba, Queensland 4350, Australia\\
$^{3}$Department of Earth Sciences, University of California, Riverside, California 92521, USA
}

\date{Accepted 2019 January 23. Received 2019 January 22; in original form 2018 October 25}

\pubyear{2017}

\begin{document}
\label{firstpage}
\pagerange{\pageref{firstpage}--\pageref{lastpage}}
\maketitle

\begin{abstract}
We present an approach that is able to both rapidly assess the dynamical stability of multiple planet systems, and determine whether an exoplanet system would be capable of hosting a dynamically stable Earth-mass companion in its habitable zone. We conduct a suite of numerical simulations using a swarm of massless test particles in the vicinity of the orbit of a massive planet, in order to develop a predictive tool which can be used to achieve these desired outcomes. In this work, we outline both the numerical methods we used to develop the tool, and demonstrate its use. We find that the test particles survive in systems either because they are unperturbed due to being so far removed from the massive planet, or due to being trapped in stable mean motion resonant orbits with the massive planet. The resulting unexcited test particle swarm produces a unique signature in ($a$,$e$) space that represents the stable regions within the system. We are able to scale and translate this stability signature, and combine several together in order to conservatively assess the dynamical stability of newly discovered multiple planet systems. We also assess the stability of a system's habitable zone and determine whether an Earth-mass companion could remain on a stable orbit, without the need for exhaustive numerical simulations.

\end{abstract}


%%%%%%%%%%%%%%%%%%%%%%%%%%%%%%%%%%%%%%%%%%%%%%%%%%

%%%%%%%%%%%%%%%%% BODY OF PAPER %%%%%%%%%%%%%%%%%%

\section{Introduction}
\label{sec:introduction}
The search for potentially habitable worlds is an area of immense interest to the exoplanetary science community. Since the first discoveries of exoplanets orbiting main sequence stars \citep{Campbell1988,Latham1989,Mayor1995}, planet search surveys have endeavoured to discover the degree to which the Solar system is unique, and to understand how common (or rare) are planets like the Earth \citep[e.g.][]{Howard2010b,Wittenmyer2011a}. The launch of the {\it Kepler} spacecraft in 2009 led to a great explosion in the number of known exoplanets \citep[e.g.][]{Borucki2011,Sullivan2015,Morton2016,Dressing2017}. {\it Kepler} carried out the first census of the Exoplanet era, discovering more than two thousand planets\footnote{As of 13 September 2018 (\textit{Kepler} and \textit{K2} mission site, \url{https://www.nasa.gov/mission_pages/kepler/main/index.html}).}. {\it Kepler}'s results offer the first insight into the true ubiquity of planets, and the frequency with which Earth-size planets within the Habitable Zone (HZ) can be found orbiting Sun-like stars \citep{Catanzarite2011,Petigura2013,Foreman-Mackey2014}. The Transiting Exoplanet Survey Satellite (\textit{TESS}) seeks to continue this trend in exoplanet discoveries \citep{Ricker2014,Sullivan2015,Barclay2018}, concentrating on planet detection around bright stars that are more amenable to follow-up spectroscopy \citep{Kempton2018}.

In addition to the rapid increase in the known exoplanet population, the generational improvement of instruments being used for detection, confirmation, and observational follow-up, has recently allowed for planets to be detected with masses comparable to that of the Earth, albeit on orbits that place them far closer to their host stars than the distance between the Earth and the Sun \citep[e.g.][]{Vogt2015,Wright2015,Anglada-Escude2016,Gillon2017}. Whilst such Earth-mass planets generate larger, more easily detectable radial velocity signals, current and near-future instruments (e.g. the ESPRESSO and CODEX spectrographs) seek to detect planets inducing Doppler wobbles as low as $0.1~\textrm{ m s}^{-1}$ and $~0.01~\textrm{m s}^{-1}$ respectively \citep{Pasquini2010,Pepe2014,Hernandez2017}. At such small detection limits, those spectrographs may well be able to detect Earth-mass planets that orbit Sun-like stars at a distance that would place them within the star's HZ \citep{Agnew2017,Agnew2018,Agnew2018b}. With the advent of these new high-precision spectrographs, as well as the launch of the next generation of space telescopes (such as the James Webb Space Telescope, {\it JWST}), it will be possible to perform observational follow-up in far greater detail. As a result, it is timely to consider methods by which we might prioritise the observations needed in order to detect Earth-size planets \citep{Horner2010}.

Whilst the detection of a Solar system analogue would be considered the holy grail in the search for an exoplanet that truly mimics Earth, finding something that resembles the Solar system itself beyond a handful of similarities has so far proven to be particularly challenging \citep{Boisse2012,Wittenmyer2014,Wittenmyer2016,Kipping2016,Rowan2016,Agnew2018}. By relaxing the criteria of our search to instead look for multiple planet systems where at least one planet is comparable in mass to Earth and resides in its host star's HZ \citep{Kasting1993,Kopparapu2013,Kopparapu2014,Kane2016}, we will still yield exoplanetary systems that share several similarities with our own system, and are still candidates for further study from the perspective of planetary habitability. Additionally, we consider the notion that the observational bias inherent to several detection methods \citep{Wittenmyer2011a,Dumusque2012b} can be interpreted to suggest that systems with massive, giant planets may also coexist with smaller, rocky exoplanets that so far have been undetectable due to detection limits \citep{Agnew2017}. Indeed, given the challenges involved in finding habitable exo-Earths, it might be the case that such planets exist within known exoplanetary systems, lurking below our current threshold for detectability. Future studies of those systems might allow such planets to be discovered, if they exist -- which serves as an additional motivation for the development of a method by which we can prioritise systems as targets for the search for Earth-like planets: systems that host dynamically stable HZs.

% As such, a habitable exo-Earth may exist within the current, or near-future, exoplanet populations that has yet been undetectable due to technological limitations. This motivates how we propose to prioritise systems in the search for Earth-like planets: systems that host dynamically stable HZs.

The standard method used to determine whether a system could host unseen planetary companions is by assessing the system's overall dynamical stability. This can be done analytically \citep[e.g.][]{Giuppone2013,Laskar2017}, or numerically \citep[e.g.][]{Raymond2005,Jones2005,Rivera2007,Jones2010,Wittenmyer2013,Agnew2017}. Such studies have resulted in the development of several methods that allow exoplanetary systems to be assessed by planetary architectures as they are observed today \citep{Giuppone2013,Carrera2016,Matsumura2016,Agnew2018}. It has been frequently demonstrated that a variety of resonant mechanisms are often integral in determining the dynamical stability of a system \citep[e.g.][]{Wittenmyer2012,Gallardo2014,Kane2015,Gallardo2016,Mills2016,Luger2017,Delisle2017,Agnew2017,Agnew2018b}. Typically, such studies examine a single exoplanetary system, and study it in some depth to determine whether it is truly dynamically stable, and whether other planets could lurk undetected within it. Such efforts are extremely computationally intensive, which means that few systems can readily be studied in such detail. 

Here, we consider whether it is instead possible to use detailed $n$-body simulations to build a more general predictive tool that could allow researchers to quickly assess the potential stability (or instability) of any given system without the need for their own suite of numerical simulations. This would identify systems for which further observational investigation might be needed. In doing so, we develop a tool by which one can: 1) rapidly and conservatively assess the dynamical stability of a newly discovered multiple planet system to identify those where further observation is required to better constrain orbital parameters, and 2) identify the regions in a given exoplanetary system that could host as-yet undiscovered planets, based on their dynamical interaction with the known planets in the system. We present a suite of simulations that have identified the stable regions around generic planetary systems that can be used to examine the stability of specific systems. Ultimately, we wish to demonstrate the use cases of our predictive tool to rapidly assess new systems found with \textit{TESS}.

In section \ref{sec:dynamics} we introduce the use cases that we have in mind when developing our predictive tool. We outline the fundamental approach of our method in section \ref{sec:method}, and demonstrate how the general stability signatures we compute can be normalised and translated to fit any single planetary system discovered. In section \ref{sec:results} we show how the mass, eccentricity and inclination of a massive body influences its stability signature. We follow this with various examples of how our method can be used to infer system stability or to determine where stable resonant HZ companions may exist in section \ref{sec:appls}, and summarise our findings in section \ref{sec:summary}.

% ----------------- DYNAMICS -----------------
\section{Dynamical Predictions}
\label{sec:dynamics}
The predictive tool we present was developed to be applied to exoplanetary systems discovered by \textit{TESS} with two goals in mind: 1) to determine the dynamical feasibility of newly discovered multiple planet systems, and 2) to predict the regions of stability in \textit{TESS} systems where another unseen planet may exist.

The first use case is to provide a tool that may be incorporated into the Exoplanet Follow-up Observing Program for \textit{TESS} (ExoFOP-TESS)\footnote{\url{https://heasarc.gsfc.nasa.gov/docs/tess/followup.html}}. When a multiple planet system is discovered, we can utilise our tool to dynamically assess the stability of the system given the inferred orbital parameters. While our approach will miss more complex destabilising behaviour \citep[such as the influence of secular resonant interactions as demonstrated by][]{Agnew2018b}, it can conservatively assess the dynamical stability of a system ``on the fly''. Demonstrating instability using a conservative approach would suggest further observations are required to better constrain the orbital parameters of the planets, or to reassess the number of planets in the system in the cases of potential eccentricity harmonics and aliasing \citep{Anglada-escud2010,Anglada-Escude2010,Wittenmyer2013}.

The second use case is to assist in the search for potentially habitable Earth-size planets. Using $n$-body simulations, we can make dynamical predictions of regions in the ($a$,$e$) parameter space where additional planets will definitely be unstable. This formed the core of our previous work in this series \citep{Agnew2017,Agnew2018,Agnew2018b}, where we were able to identify systems that could potentially host dynamically stable HZ planets. Here, we extend that work to develop a stability mapping process that will enable future studies to quickly identify the regions in newly discovered exoplanetary systems where planets can definitely be ruled out, on dynamical grounds, as well as those regions where additional planets could only exist under very specific conditions \citep[such as when trapped in a mutual mean-motion resonance with another planet, e.g.][]{W.Howard2010a,Robertson2012,Wittenmyer2014b}. 

To develop such a tool, we performed a large number of detailed $n$-body simulations to determine how the unstable regions centred on a given planet's orbit are influenced by its mass, orbital eccentricity and inclination in order to produce a scale free template that can be applied to any system.

% ----------------- METHOD ----------------- 
\section{Method}
\label{sec:method}
We have developed our predictive tool by conducting thorough, high resolution test particle (TP) simulations. In each simulation, a swarm of TPs are distributed randomly throughout a region in ($a$,$e$) space around the orbit of a massive planet. At the end of the simulation, the distribution of surviving TPs reveals evacuated regions (which correspond to areas of instability), and regions where TPs remain unperturbed, and so remain on similar orbits to those held at the beginning of the integrations. Since our simulations are primarily focused on facilitating the search for habitable worlds, we chose to enforce a maximum initial eccentricity on the orbits of TPs of 0.3, based on studies that suggest that the habitability of a given Earth-like world would be significantly reduced for higher eccentricities \citep{Williams2002,Jones2005}.

The goal of these simulations is to determine the stable, unperturbed regions around a massive body, which we refer to as that body's ``stability signature''. More specifically, we put forward two definitions: the optimistic stability signature being the curve that bounds the maximum eccentricity of the unexcited TPs, and the conservative stability signature that bounds the minimum eccentricity of the excited TPs. The key reasons for considering the signature as a curve rather than a 2-dimensional area in ($a$,$e$) space are: 1) in the simplest case (the massive body moving on a circular orbit) only those TPs with sufficiently eccentric orbits have apsides that enter within the region of instability nearby to the massive planet (generally some multiple of its Hill radius), and so the regions below the curve (i.e. the TPs with less eccentric orbits) will be stable, and 2) multiple curves can be combined and presented in a ``look-up map'' within which we can interpolate to find the curves for planetary masses we do not simulate explicitly. This is explained in greater detail in section~\ref{sec:results}.

By determining the stability signatures for a range of planet masses, the signatures can be used as a scale free template that can be applied to any system in order to assess its dynamical stability without the need to run numerical simulations. Our proposed exploration of the planet mass parameter space and the resulting stability signatures will also enable researchers to predict the regions of a given exoplanetary system where additional planets are dynamically feasible, and those where no planets are likely to be found. We anticipate that this method will provide a useful filtering tool by which systems can be examined to quickly predict whether they are dynamicaly feasible, and which might be able to host a potentially habitable exoplanet. We present here the general simulations we carried out from which we determined the stability signatures and how the signatures from these simulations can be used to create the stability template for any specific system.

In this work, we consider the stability of additional bodies moving on co-planar orbits in circular, single planet systems. While it is highly unlikely that a planet will have a perfectly circular orbit, our intent is for this to be a conservative assessment of the stability of the system. Our approach can be expanded and refined to also consider eccentric and inclined orbits, which we will pursue in future work.

\subsection{Numerical approach}
\label{subsec:fund}
Two observables when detecting an exoplanet are stellar mass, $M_\star$ (obtained indirectly from the luminosity of the star, $L_\star$), and the orbital period of a planet, $T_\textrm{pl}$ (obtained directly). We can use these to calculate the semi-major axis of the planet by
\begin{align}\label{eqn:gravity}
	a_\textrm{pl} = \sqrt[3]{\frac{GM_\star T_\textrm{pl}^2}{4\pi^2}},
\end{align}
where $G$ is the universal Gravitational constant, and hence we can infer the distance of a planet from the observed orbital period. For any fixed $M_\star$ value, the semi-major axis will scale with orbital period according to the power law
\begin{align}
	a \propto T^{2/3}.
		\label{eq:a_scale_T}
\end{align}

\cite{Kopparapu2014} put forward a method to calculate the HZ of a system using $M_\star$ and the planetary mass, $M_\textrm{pl}$. In general, a minimum mass for $M_\textrm{pl}$ is determined via the radial velocity method. Taken in concert with the above, this means that from the observed parameters $L_\star$ and $T_\textrm{pl}$, we can infer $M_\star$ and $a_\textrm{pl}$, measure $M_\textrm{pl}$, and hence compute the HZ around the star.

By numerically simulating a range of mass ratios ($\mu = M_\textrm{pl}/M_\star$) while keeping constant the mass of the star ($M_\star=\mathrm{M}_\odot$) and semi-major axis of the planet ($a_\mathrm{pl}=1$~au), we produce the stability signature for each of the simulated masses and can consolidate them to make a ``look-up map''. 
% For a newly discovered system, we can first normalise it to a $\mathrm{M}_\odot$ system, adjusting the planet mass to keep the mass ratio constant, and then interpolate between our simulated stability signatures in our look-up map.
To apply our general stability signatures to a newly discovered system, one would first determine the mass ratio between the planet and the star in that system, and use that mass ratio to interpolate between our simulated stability signatures in our look-up map. This yields the signature around a planet of any mass $M_\textrm{pl}$. Once this step is complete, one then translates the signature obtained from the nominal $a_\mathrm{pl}=1$~au to a location closer to, or farther from, the host star to match the discovered planet. We can then compare the stability signatures of all planets in multiple planet systems with one another to conservatively assess dynamical stability, as well as determine which systems can host hypothetical exo-Earths within their HZ, without the need to run numerical simulations.

This approach limits the stability constraint to be multiples of the orbital period rather than number of years. Our numerical simulations are for a planet orbiting at $a_\textrm{pl} = 1$ au in a $M_\star = 1~\mathrm{M}_\odot$ system (i.e. $T_\textrm{pl} = 1$ year). As we simulate our systems for $10^7$ years (i.e. $10^7$ orbits), if we were to use our stability signature for a planet that orbits at a semi-major axis with an orbital period of $0.1$ years (i.e. $1/10^{\textrm{th}}$ of the simulated orbital period) we can only refer to the planet as being stable for $10^6$ years (i.e. $10^7$ orbits).

\begin{figure}
	\begin{subfigure}{\linewidth}
    		\centering
    		\includegraphics[width=\linewidth]{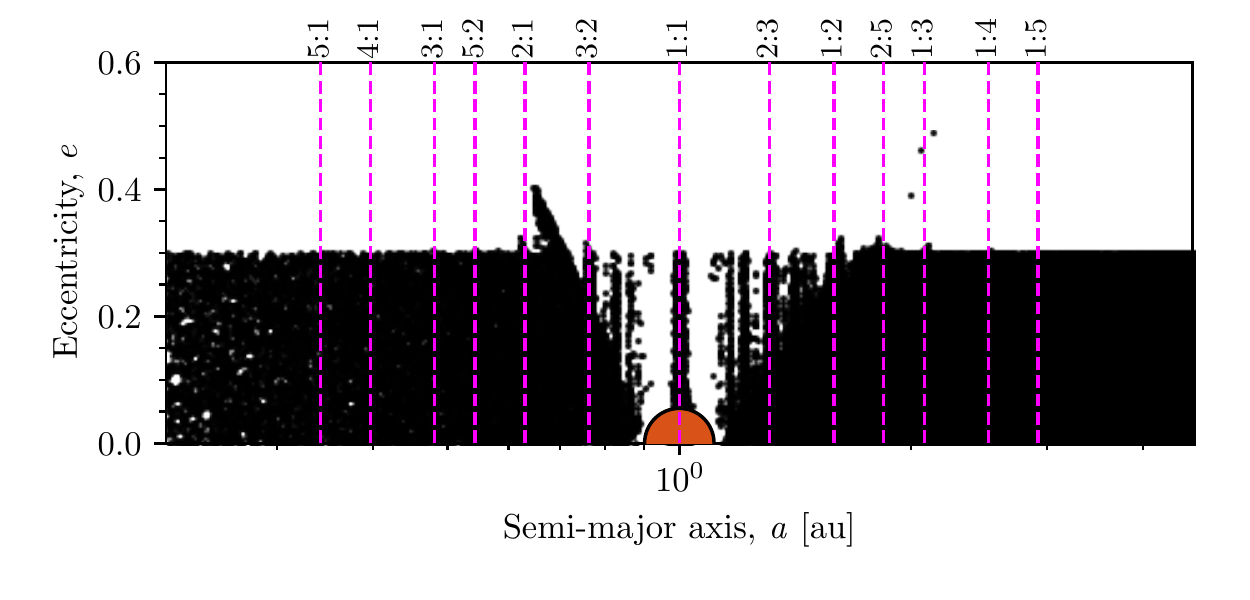}
        \caption{$\mu={64 \textrm{M}_{\oplus}}/\textrm{M}_\odot$ at $1$~au}
        \label{fig:normalise_before}
	\end{subfigure}
	
    \begin{subfigure}{\linewidth}
	\centering
		\includegraphics[width=\linewidth]{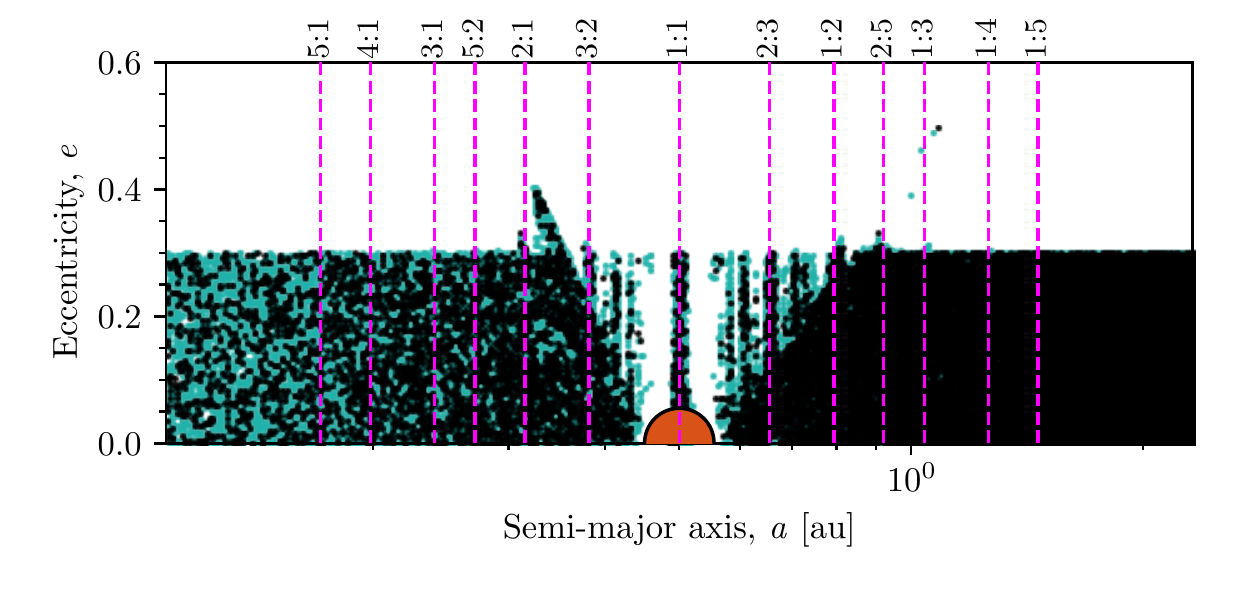}
		\caption{$\mu={8 \textrm{M}_{\oplus}}/(0.25\textrm{M}_\odot)$ at $0.5$~au}
		\label{fig:normalise_near}
	\end{subfigure}
	
	\begin{subfigure}{\linewidth}
	\centering
		\includegraphics[width=\linewidth]{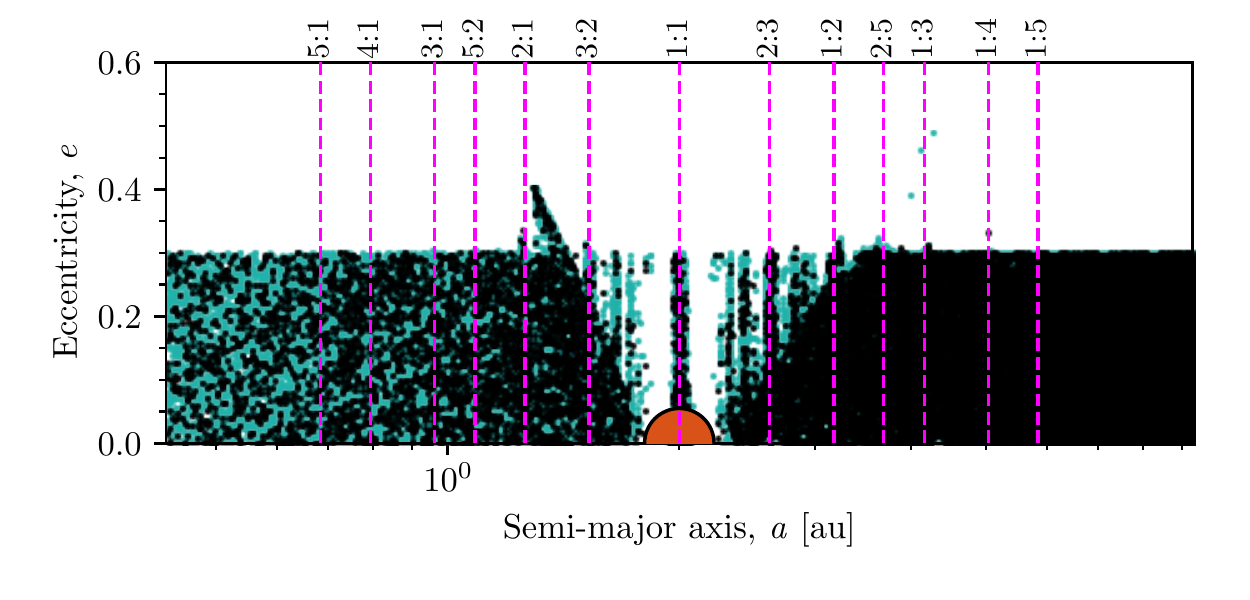}
		\caption{$\mu={512 \textrm{M}_{\oplus}}/(8\textrm{M}_\odot)$ at $2$~au}
		\label{fig:normalise_far}
	\end{subfigure}
	\caption{A demonstration of how normalisation of a stability signature is possible. The black points represent those test particles that have not been removed from the system at the end of the simulation. We plot the points from (a) as teal points in (b) and (c) to more easily compare structure between the simulations with different $M_\mathrm{pl}$. The mass ratio remains constant ($\mu={64 \textrm{M}_{\oplus}}/\textrm{M}_\odot$) in all three simulations. By ensuring the semi-major axis of the planet is such that its orbital period remains constant, i.e. $T_{\mathrm{pl}}=1~\mathrm{year}$ at (a) $1$~au around $1~\textrm{M}_\odot$, (b) $0.5$~au around $0.25~\textrm{M}_\odot$ and (c) $2$~au around $8~\textrm{M}_\odot$, then the stability signature across differing mass planets remains constant, as shown by the agreement between the black and teal points. Some common low order MMRs are shown in pink.}\label{fig:normalisation}
\end{figure}

\subsection{Normalisation of system}
\label{subsec:norm}
As our general simulations were performed for a range of mass ratios, but for a fixed stellar mass, normalisation is important so that we are still able to determine the stability signature around a newly discovered system regardless of its stellar mass. 

As per our definition of the stability signature being the curve that bounds the maximum eccentricity of the unexcited TPs, let us consider a hypothetical function that maps semi-major axis, $a$, to this curve, $\{(a, f(a)):a\in A\}$, where $A$ is the domain over which we simulated. Our simulations all use a stellar mass of $M_\star = 1~\mathrm{M}_\odot$, and a planet orbiting at $a_\mathrm{pl}=1$~au (i.e. $T_\mathrm{pl}=1$~yr). To normalise a system, we first determine what the semi-major axis is that matches an orbital period of one year for a system with a different stellar mass. From equation~\ref{eqn:gravity}, the semi-major axis will scale with stellar mass according to the power law
\begin{align}
	a \propto M_\star^{1/3}
		\label{eq:a_scale_M}.
\end{align}
For a given system discovered with stellar mass $M_\star=n \mathrm{M}_\odot$, we can scale the masses $M_\star$ and $M_\mathrm{pl}$ by $1/n$, and compute that the one year orbital period will occur at $a_\mathrm{1year} = n^{1/3}$~au. We demonstrate in Figure~\ref{fig:normalisation} how the stable signatures are shown to be identical when the mass ratio, $\mu$, and orbital period, $T_\textrm{pl}$, remain constant. Thus, when an exoplanet of mass $M_\mathrm{pl}$ is discovered orbiting a star of mass $M_\star$, it is possible to first normalise the system by scaling the planetary and stellar masses while retaining a constant mass ratio.

Once normalised, we can then interpolate between the planetary masses we simulated in order to obtain the stability signature for any planetary mass (within the upper and lower $\mu$ we simulate). In terms of function notation, a new function for the normalised stability signature $f_\mathrm{norm}$ will have the scaled domain of $f$, while still mapping to the unscaled stability signature values $f(a)$, that is,
\begin{align}
    \{(a, f_\mathrm{norm}(a)):a\in n^{1/3} A\},
\end{align}
where $f_\mathrm{norm}(a)$ is the function mapping the scaled domain to the original signature curve, given by
\begin{align}
    f_\mathrm{norm}(n^{1/3}a) = f(a). \nonumber
\end{align}
We must then determine how the domain will vary for a function that is translated from the semi-major axis where an orbital period of one year occurs, $a_\mathrm{1year}$, to the semi-major axis of the discovered exoplanet, $a_\mathrm{pl}$.

\subsection{Translatability of signature}
\label{subsec:trans}
As our general simulations are run for a planet at a fixed semi-major axis of $a_\mathrm{pl} = 1~\mathrm{au}$, translatability is important so that we are still able to determine the stability signature around a newly discovered system regardless of the semi-major axis of the planet. We translate the normalised signature by simply taking the ratio between the semi-major axis of the planet ($a_\mathrm{pl}$) and the one year semi-major axis ($a_\mathrm{1year}$). We express this ratio as
\begin{align}\label{eqn:a_scale}
	k &= {a_\mathrm{pl}}/{a_\mathrm{1year}} \nonumber\\
	&= {a_\mathrm{pl}}/n^{1/3}.
\end{align}
The discovered exoplanet's stability signature at its semi-major axis is then obtained by multiplying the normalised signature by $k$. In terms of function notation, a new function for the planet's stability signature $f_\mathrm{planet}$ will have the translated domain of $f_\mathrm{norm}$, while still mapping to the unscaled stability signature values $f(a)$, that is,
\begin{align}
    \{(a, f_\mathrm{planet}(a))&:a\in k\ n^{1/3}\ A\}\nonumber\\
    \{(a, f_\mathrm{planet}(a))&:a\in ({a_\mathrm{pl}}/n^{1/3})\ n^{1/3}\ A\}\nonumber\\
    \{(a, f_\mathrm{planet}(a))&:a\in a_\mathrm{pl}\ A\}
\end{align}\label{eq:translate}
where $f_\mathrm{planet}(a)$ is the function mapping the scaled and translated domain to the original signature curve, that is,
\begin{align}
    f_\mathrm{planet}(a_\mathrm{pl}\ a) = f(a). \nonumber
\end{align}
Thus, from the observed parameters $L_\star$ and $T_\textrm{pl}$, we can infer $M_\star$ and $a_\textrm{pl}$, and from radial velocity measurements we can determine $M_\textrm{pl}$. With these parameters we are able to: 1) normalise the system by scaling  $M_\star$ and $M_\textrm{pl}$, 2) interpolate between the planetary masses in our look-up map in order to obtain the normalised signature, and 3) translate the normalised signature to the semi-major axis of the detected planet to find its stability signature \textit{without the need for additional numerical simulations}. Being able to obtain the stability signature of any discovered planetary system allows us to: 1) combine the stability signatures of all planets within a multiple planet system in order to rapidly and conservatively assess its dynamical stability, and 2) determine the stable, unperturbed regions around an exoplanet in order to constrain where a habitable terrestrial planet could exist. These applications will be discussed in depth in section~\ref{sec:appls}.

\begin{figure}
	\begin{subfigure}{\linewidth}
    		\centering
    		\includegraphics[width=\linewidth]{images/normalisation/before_after_a_m64.pdf}
        \caption{$\mu={64 \textrm{M}_{\oplus}}/\textrm{M}_\odot$ at $1$~au}
        \label{fig:translate_before}
	\end{subfigure}
	
    \begin{subfigure}{\linewidth}
	\centering
		\includegraphics[width=\linewidth]{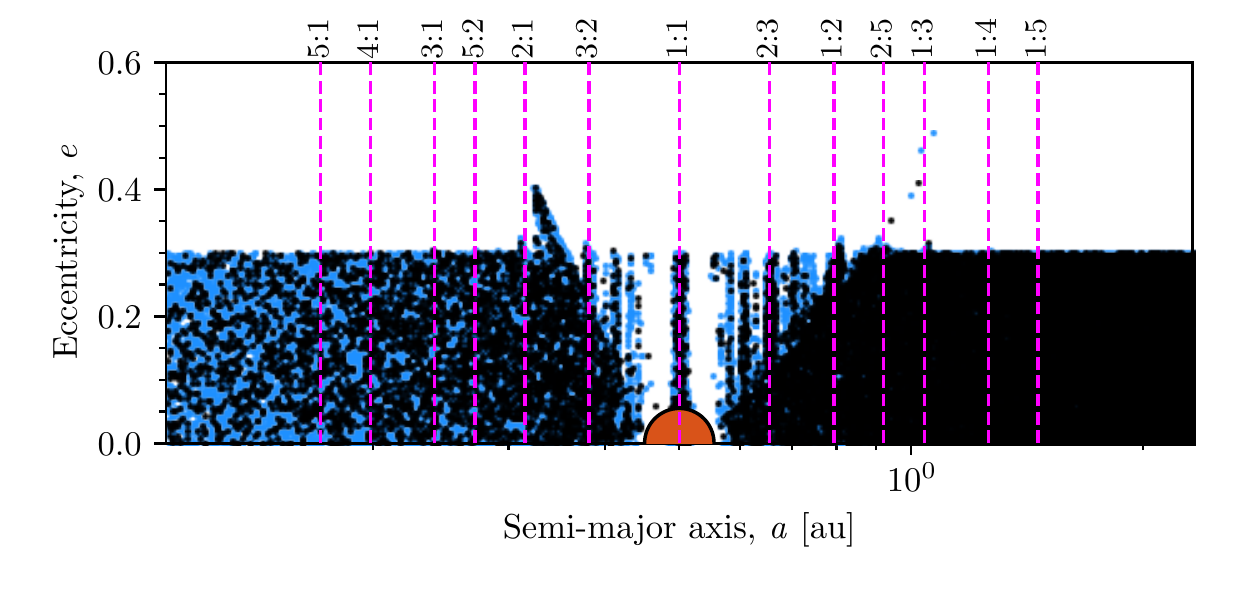}
		\caption{$\mu={64 \textrm{M}_{\oplus}}/\textrm{M}_\odot$ at $0.5$~au}
		\label{fig:translate_near}
	\end{subfigure}
	
	\begin{subfigure}{\linewidth}
	\centering
		\includegraphics[width=\linewidth]{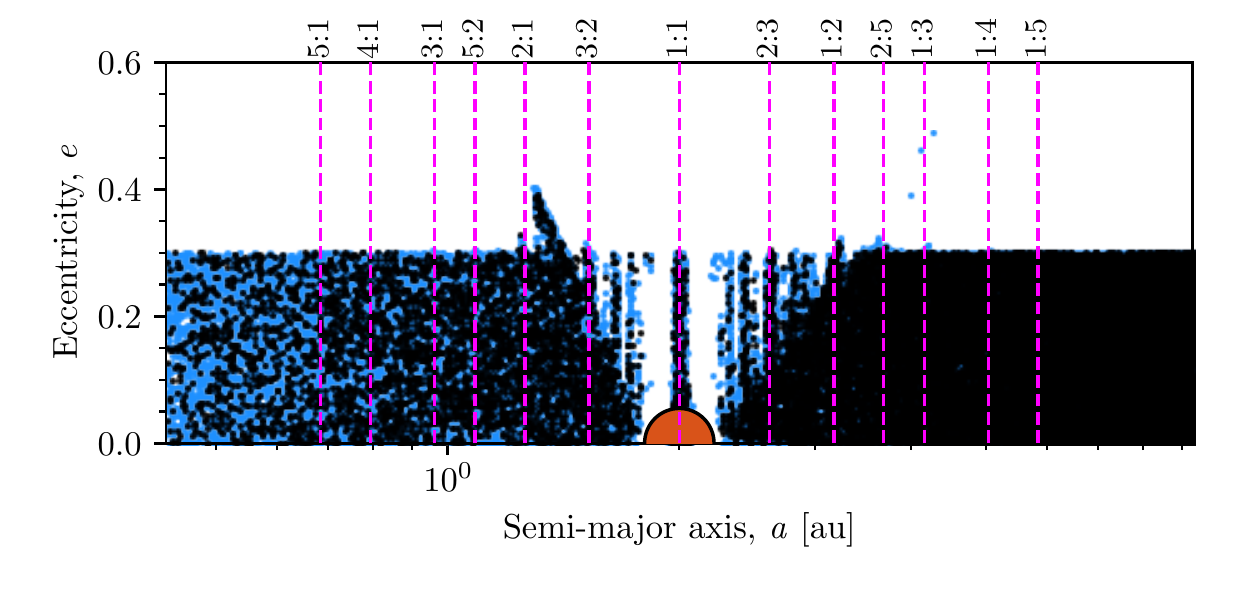}
		\caption{$\mu={64 \textrm{M}_{\oplus}}/\textrm{M}_\odot$ at $2$~au}
		\label{fig:translate_far}
	\end{subfigure}
\caption{A demonstration of how translatability of a signature is possible. The black points represent those test particles that have not been removed from the system at the end of the simulation. We plot the points from (a) as blue points in (b) and (c) to more easily compare structure between the simulations. The mass ratio remains constant ($\mu={64 \textrm{M}_{\oplus}}/\textrm{M}_\odot$) in all three simulations. The stability signature is then translated to different semi-major axes i.e. from (a) $\textrm{a}_{\mathrm{pl}}=1~\mathrm{au}$ to (b)  $\textrm{a}_{\mathrm{pl}}=0.5~\mathrm{au}$ or (c)  $\textrm{a}_{\mathrm{pl}}=2~\mathrm{au}$. However, as outlined in section~\ref{sec:method}, the number of years of stability is determined by the semi-major axis the signal has been translated to. Some common low order MMRs are shown in pink.}\label{fig:translatability}
\end{figure}
\subsection{Simulations}
\label{subsec:simulations}
All our simulations model a two-body, star--planet system within which we randomly scatter $10^5$ massless TPs in order to test the stability of a hypothetical third body. The motion of two massive bodies and a third massless body constitutes the Restricted Three-Body Problem. The $10^5$ TPs represent $10^5$ possible orbital parameter configurations for this third body. These TPs are scattered in order to find the stable regions within the ($a$,$e$) parameter space within the vicinity of the orbit of a massive body. The central star for our simulations has mass $M_\star = 1~\mathrm{M}_\odot$, and the semi-major axis of the planet is $a_\mathrm{pl} = 1~\mathrm{au}$. We then carry out three distinct suites of simulations, in which we vary the planetary mass ($m_\mathrm{pl}$), eccentricity ($e_\mathrm{pl}$) or inclination ($i_\mathrm{pl}$). The orbital parameters of the planet for each suite of simulations is summarised in Table~\ref{tab:pl_params}. The orbital parameters for the swarm of TPs are randomly generated within a fixed range for all three sets of simulations, and will be discussed in detail after first outlining how the planetary parameters are varied in each simulation.
\begin{table}
	\caption{The orbital parameters of the massive planet used in each set of simulations.}
   	\label{tab:pl_params}
   	\centering
	\begin{tabular}{c c c c}
    	\toprule                
        Parameter				    & Set 1	& Set 2  & Set 3\\
        \midrule
        $M$ ($\mathrm{M}_\oplus$)   & $1,2,4, ... ,1024$& 32.0  & 32.0 \\
        $a$ (au)				    & 1.0   & 1.0   & 1.0 \\
        $e$					        & 0.0	& $0, 0.05, 0.1, ... , 0.3$& 0.0 \\
        $i$ ($\degree$)		        & 0.0	& 0.0   & $0,2.5,5, ... , 10$ \\
        $\Omega$ ($\degree$)	    & 0.0	& 0.0   & 0.0 \\
        $\omega$ ($\degree$)	    & 0.0	& 0.0   & 0.0 \\
        $M$ ($\degree$)	        	& 0.0	& 0.0   & 0.0 \\
        \bottomrule 
   	\end{tabular}
\end{table}
In our first set of simulations, we explore the effect of planetary mass, $m_\mathrm{pl}$. Since the mass of the host star is kept constant, this allows us to examine the effect of the mass ratio $\mu = M_\mathrm{pl}/M_\star$ on the resulting stability signatures. In these simulations, we set the orbital parameters of the planet to those values shown for Set 1 in Table~\ref{tab:pl_params}. The mass of the planet is then varied sampling the range $M_\mathrm{pl} = 1$ to $1024\ \textrm{M}_{\oplus}$, with the mass increased incrementally by factors of 2 (i.e. $1, 2, 4, 8, ...$). We span these masses so as to appropriately cover the expected mass distribution of planets found with \textit{TESS} \citep{Sullivan2015}.
% , as shown in Table~\ref{tab:exoplanet_distribution}.

% \begin{table}
%     \caption{The expected distribution of planets forecast to be found with \textit{TESS} \protect\citep{Sullivan2015}, and the corresponding radius and mass of each.}
%     \label{tab:exoplanet_distribution}
%     \centering
%     \begin{tabular}{l c c c c c}
%         \toprule
%         					& $r_{\mathrm{min}}$ 		& $r_{\mathrm{max}}$  		& $m_{\mathrm{min}}$  		& $m_{\mathrm{max}}$ 	  	& Occurrence\\                     
%         					& ($\textrm{r}_{\oplus}$) 	& ($\textrm{r}_{\oplus}$) 	& ($\textrm{M}_{\oplus}$) 	& ($\textrm{M}_{\oplus}$) 	& ($\%$)\\
%         \midrule
%         Terrestrials		& -							& $<1.25$					& -							& $<5$ 						& 4\\
%         Super-Earths		& 1.25						& $<2$						& 5							& $<10$ 						& 28\\
%         Neptunians		& 2							& $<4$						& 10							& $<50$ 						& 64\\
%         Jovians			& 4							& $>4$						& 50							& $>50$ 						& 4\\
%         	\midrule  
%     \end{tabular}
% \end{table}

In our second set of simulations we explore the effect of eccentricity. In these simulations we fix the planetary mass to $M_\mathrm{pl} = 32\ \textrm{M}_{\oplus}$ as this falls within the dominant mass range of expected \textit{TESS} findings \citep{Sullivan2015}. In these simulations, we set the orbital parameters of the planet to those values shown for Set 2 in Table~\ref{tab:pl_params}. The eccentricity of the planet is then varied from $e_\mathrm{pl} = 0.0$ to $0.3$ in steps of $0.05$ for each simulation.
% (\textit{Neptunians} in Table~\ref{tab:exoplanet_distribution}).
In our final set of simulations, we explore the effect of inclination. In these simulations we again fix the planetary mass to $M_\mathrm{pl} = 32\ \textrm{M}_{\oplus}$. and set the orbital parameters of the planet to those values shown for Set 3 in Table~\ref{tab:pl_params}. The inclination of the planet is then varied from $i_\mathrm{pl} = 0.0\degree$ to $10.0\degree$ in steps of $2.5\degree$ for each simulation.

For all of our simulations we randomly scatter $10^5$ massless TPs throughout the system. As we are interested in obtaining the stability signatures in the vicinity of the orbit of a massive body, the TPs are distributed between orbits with periods in $10:1$ and $1:10$ commensurability with the planet (which for $a_\mathrm{pl} = 1~\mathrm{au}$ is given by $0.215~\mathrm{au}\lesssim a\lesssim4.642~\mathrm{au}$), and with eccentricities $0.0 \leq e \leq 0.3$. It should also be noted that, by varying the values of $\omega_\mathrm{tp}$ and $M_\mathrm{tp}$ across the full $0$ to $360\degree$ range, this allows for a fixed value of $\omega_\mathrm{pl}=0.0\degree$ and $\Omega_\mathrm{pl} = 0.0\degree$ for sets 2 and 3 respectively to still cover the relevant parameter space.
% \begin{table}
% 	\caption{The orbital parameters of the TPs for the simulations. The TPs were randomly distributed between the minimum and maximum values given. }
%   	\label{tab:tp_params}
%   	\centering
% 	\begin{tabular}{c c c}
%     	\toprule                
%         						&	Min					& Max\\
%         \midrule
%         $a$ (au)				& $a$ at $1/10\ T_{\mathrm{pl}}$	& $a$ at $10\ T_{\mathrm{pl}}$\\
%         $e$					&	0.0					& 0.3\\
%         $i$ ($\degree$)		&	0.0					& 0.0\\
%         $\Omega$ ($\degree$)	&	0.0					& 0.0\\
%         $\omega$ ($\degree$)	& 	0.0					& 360.0\\
%         $M$ ($\degree$)		&	0.0					& 360.0\\
%         \bottomrule 
%   	\end{tabular}
% \end{table}

\begin{figure*}
	\begin{subfigure}{\linewidth}
		\centering
    	\includegraphics[width=\linewidth]{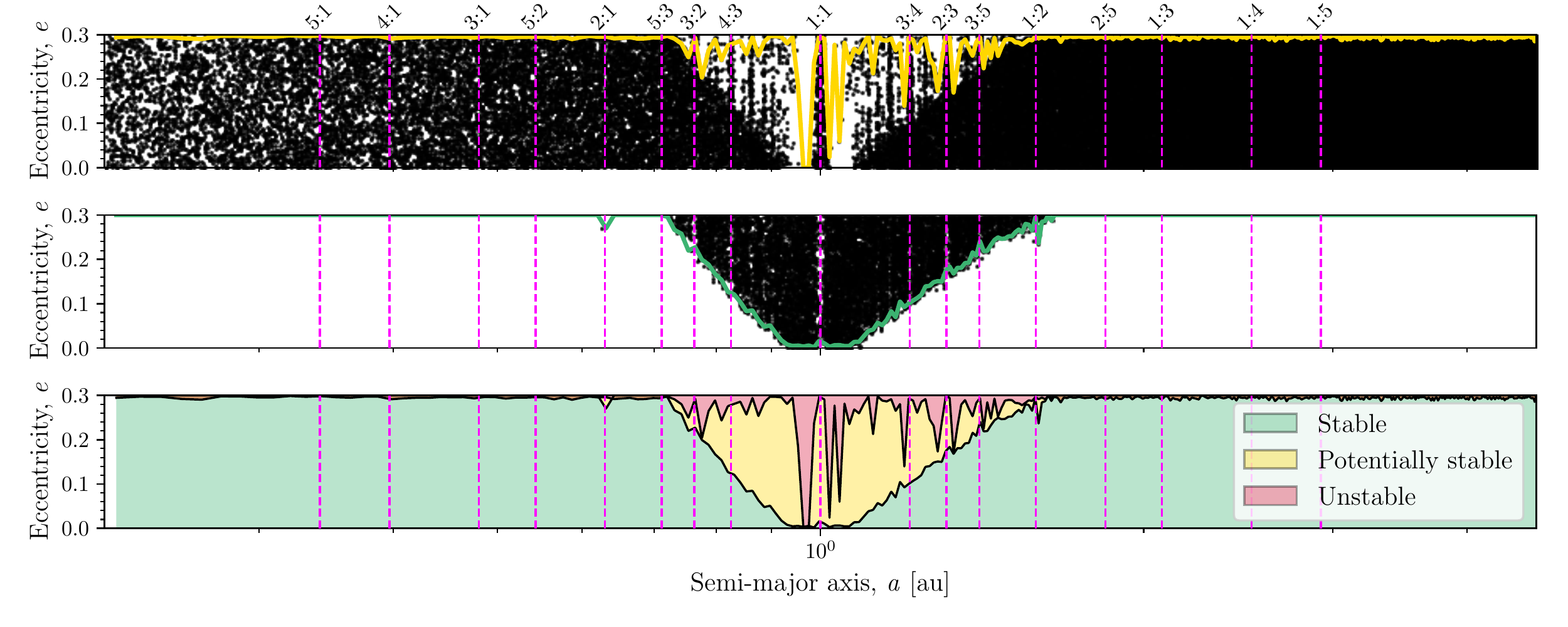}
    	\caption{Signal extraction of $\mu={8 \textrm{M}_{\oplus}}/\textrm{M}_\odot$ system}\label{fig:m8_extract}
    \end{subfigure}

	\begin{subfigure}{\linewidth}
		\centering
    	\includegraphics[width=\linewidth]{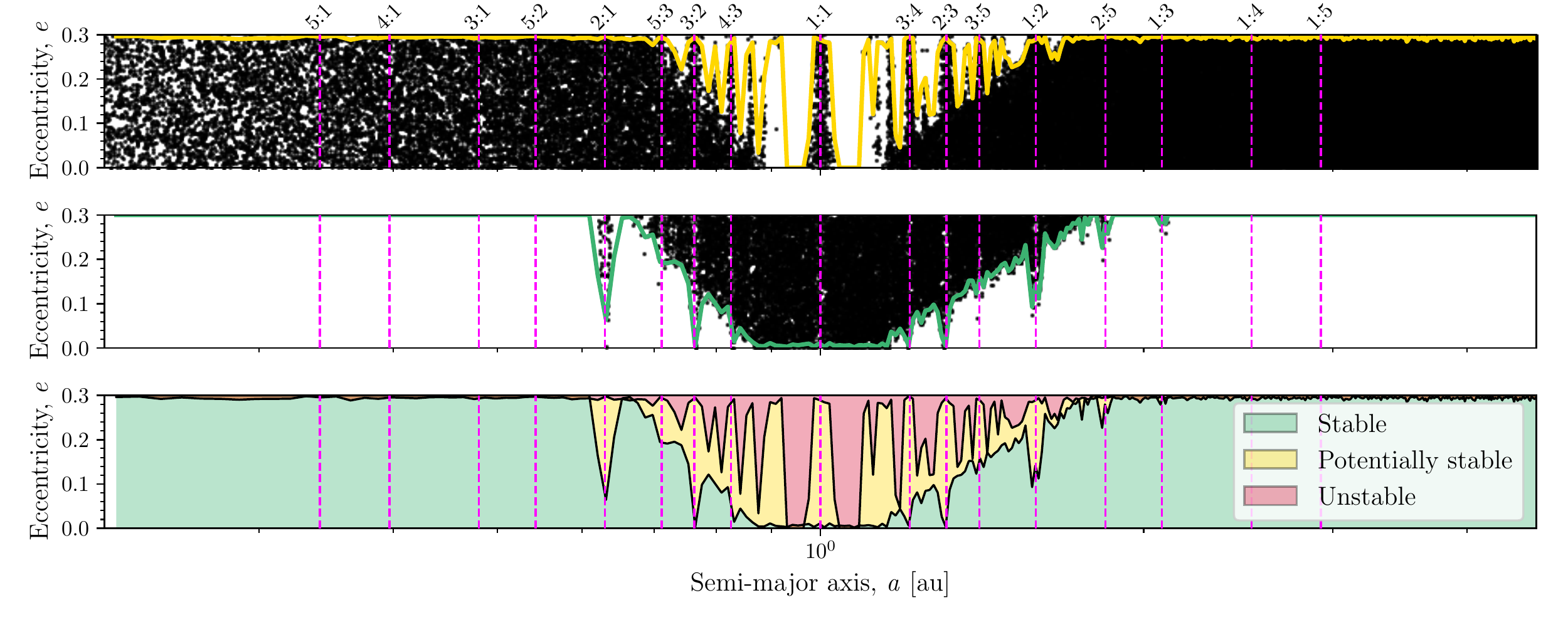}
    	\caption{Signal extraction of $\mu={64 \textrm{M}_{\oplus}}/\textrm{M}_\odot$ system}\label{fig:m64_extract}
	\end{subfigure}
	\caption{Stability signature extraction from two different systems. The TPs plotted to extract the optimistic signature (yellow) are those considered to be ``unexcited'' ($\Delta a_\mathrm{excited} < 0.01$). Conversely, the TPs plotted to extract the conservative signature (green) are those considered to be ``excited'' ($\Delta a_\mathrm{excited} > 0.01$). The semi-major axis is divided into 400 equally spaced bins, and the value of that bin is determined to be the maximum eccentricity of the unexcited TPs within that bin (for the yellow optimistic case) or the minimum eccentricity of the excited TPs within that bin (for the green conservative case). The signatures can be combined to yield a tri-colour plot with three distinct regions: unstable, stable and potentially stable. Some common low order MMRs are shown in pink.}\label{fig:signal_extract}
\end{figure*}

\subsection{Analysis}
One of the key goals of this work was the development of a tool that can be used to assess the dynamical stability of a system without the need to run numerical simulations. We achieve this by running \textit{n}-Body simulations with a swarm of massless TPs to produce a set of scalable templates, which reveals the unperturbed regions around a planet, which we refer to as that planet's ``stability signature''. More specifically, we determine two stability signatures: one that optimistically provides an upper bound above which TPs are shown to be excited, and a second that more conservatively provides a lower bound below which no TPs were shown to be excited.

For each simulation, we determine the system's stability signatures based on the excitation of the TPs perturbed by the planet over a period of $10^7$ years. We consider a TP to be excited by the planet if its semi-major axis changes relative to its initial position by $\Delta a_\mathrm{excited} = (a_\textrm{final} - a_\textrm{initial})/a_\textrm{initial} > 0.01$ over the course of the entire simulation. This is because ultimately we wish to use the signatures to predict the locations most likely to host additional planets in the system -- particularly Earth-mass bodies. Since our simulations use massless TPs, they do not take into account mutual gravitational interactions between the planet and any companion -- represented by the TPs -- orbit. In order to use TPs to predict the stability of massive bodies with any confidence \citep[such as demonstrated by ][]{Agnew2017,Agnew2018b}, we only need to consider those TPs that are not excited (in other words, the gravitational influence of the known exoplanet is not destabilising it). The gravitational influence of the less massive Earth-mass planet is also unlikely to perturb the known exoplanet in this scenario. In the cases of our first and third sets of simulations, where the massive planet moves on a circular orbit, the Jacobi integral is constant as the system is a circular restricted three-body problem. The dynamics in this case differs from the restricted three-body problem that we see with eccentric orbits, and so the results of Set 2 will also differ significantly due to the difference in the dynamical evolution resulting from the additional eccentricity of the planet itself.

Thus, by considering the relative change in semi-major axis of TPs, and ignoring those that are excited by the gravitational influence of the planet (i.e. $\Delta a_\mathrm{excited} > 0.01$), we are left with the most stable, unexcited bodies. From this, we can extract the optimistic stability signature by placing the unexcited TPs into bins in semi-major axis, and taking the maximum eccentricty of the TPs in each bin. In contrast, we obtain the conservative stability signature, below which we see no TP excitation, by placing the excited TPs into bins in semi-major axis, and taking the minimum eccentricity of the TPs in each bin. In both scenarios, we disregard outliers. We do this for two reasons. Firstly, tiny regions of stability seem to be unlikely places for planets to form and survive in all but the youngest stellar systems, and so such regions typically offer little prospect of predicting potential stable planet locations. In addition, we are interested in regions of stability that are represented by stable swarms of TPs, rather than individual outliers, as stable resonances are not infinitely thin, but have a measurable width in semi-major axis space. In this work, we ignore surviving TPs that are alone in a given bin. For bins that contain multiple surviving TPs, we consider the cumulative distribution function (CDF) of the eccentricities of the TPs. We ignore any TPs that have a final eccentricity in the top $2.5\%$ of the CDF within each bin. The converse process is true for ignoring outliers when determining the conservative stability signature, considering the excited TPs instead of the surviving TPs.

% we ignore any TPs that have a final eccentricity more than $2\sigma$ beyond the mean of all surviving TPs in that bin. The result of this cut is that we ignore the $\sim2.5\%$ of TPs that lie at the maximum and minimum ends of each bin.

Plotting a curve connecting the eccentricity of the most eccentric, unexcited TP in each semi-major axis bin -- and for which outliers have been removed -- yields a signature like that shown by the yellow curve in the upper plots of Figure~\ref{fig:signal_extract}. Similarly, we can plot the curve of the least eccentric, excited TPs as shown by the green curve (middle plots). Figure~\ref{fig:m8_extract} shows that at low masses ($\mu={8 \textrm{M}_{\oplus}}/\textrm{M}_\odot$), other than the very obvious region that is cleared in the planet's immediate vicinity, the optimistic signature is very difficult to extract, and as a result, the accuracy with which this can be achieved is somewhat limited. However, the curve of the excited TPs is much more clearly defined, and so this can still be acquired easily. Figure~\ref{fig:m64_extract} shows that for more massive planets ($\mu={64 \textrm{M}_{\oplus}}/\textrm{M}_\odot$), the outlined method for obtaining the optimistic stability signatures is much cleaner as the signature is more distinct. The optimistic stability signatures (the yellow curves in Figure~\ref{fig:signal_extract}) are thus the curve in ($a$,$e$) space, above which are the perturbed regions near a given planet. The conservative stability signatures (the green curves) are the curve in ($a$,$e$) space below which are the stable, unperturbed regions near a given planet. These can be combined as shown in the tri-colour figures (bottom plots) to yield a classification scheme where planets can be deemed stable, unstable, or potentially stable in between these two extremes where we have evidence of TPs being both excited and unexcited.

Using the approach outlined above, the stability signatures were extracted for all of our simulations and compiled to produce the desired look-up maps, and to compare how different parameters might impact dynamical stability of the systems. The resulting look-up map forms the basis of the predictive tool we developed in assessing system stability without further numerical simulations. 

% ----------------- RESULTS ----------------- 
\section{Results and Discussion}
\label{sec:results}
\subsection{Effect of mass}
The influence of the mass of the known planet on the stability in a given system is the main focus of this work, and as such is covered in the most detail. We present the compiled stability signatures and look-up maps in Figure~\ref{fig:mass_inf}.

Figure~\ref{fig:all_signals} shows the stability signature for all the planetary masses we simulated in Set 1 in Table~\ref{tab:pl_params}. The stability signatures were extracted as outlined above, and specifically the signatures for $\mu={8 \textrm{M}_{\oplus}}/\textrm{M}_\odot$ and $\mu={64 \textrm{M}_{\oplus}}/\textrm{M}_\odot$ can be seen in Figure~\ref{fig:signal_extract}. The $x$-axis shows the semi-major axes between orbits with periods in $10:1$ and $1:10$ commensurability with the planet (which for $a_\mathrm{pl} = 1~\mathrm{au}$ is given by $0.215~\mathrm{au}\lesssim a\lesssim4.642~\mathrm{au}$), while the $y$-axis shows the eccentricity values of the stability signatures in Figure~\ref{fig:all_signals}, and indicates the mass ratio of the planet in each simulation in Figures~\ref{fig:contour} and \ref{fig:contour_r}. As the maximum eccentricity we tested with our simulations was $0.3$, this means that the maximum value in these signatures correspond with very stable, unexcited regions for a body up to an eccentricity of $0.3$. While the maximum values could mean bodies with eccentricities higher than $0.3$ may also be stable at these locations, as we did not explore higher than $0.3$ we can only predict up to this value. We also overlay various other boundaries for the onset of chaos. These include simpler approximations, such as the boundaries of $1$, $3$ and $5$ Hill radii from the massive planet, as well as more developed analytic definitions. There have been several studies into the onset of chaos \citep[e.g.][]{Mustill2012,Giuppone2013,Deck2013,Petit2017,Hadden2018}, and here we plot those results presented by \cite{Mustill2012}, \cite{Giuppone2013} and \cite{Deck2013}.

\begin{figure*}
	\begin{subfigure}{\textwidth}
		\centering
    	\includegraphics[width=\linewidth]{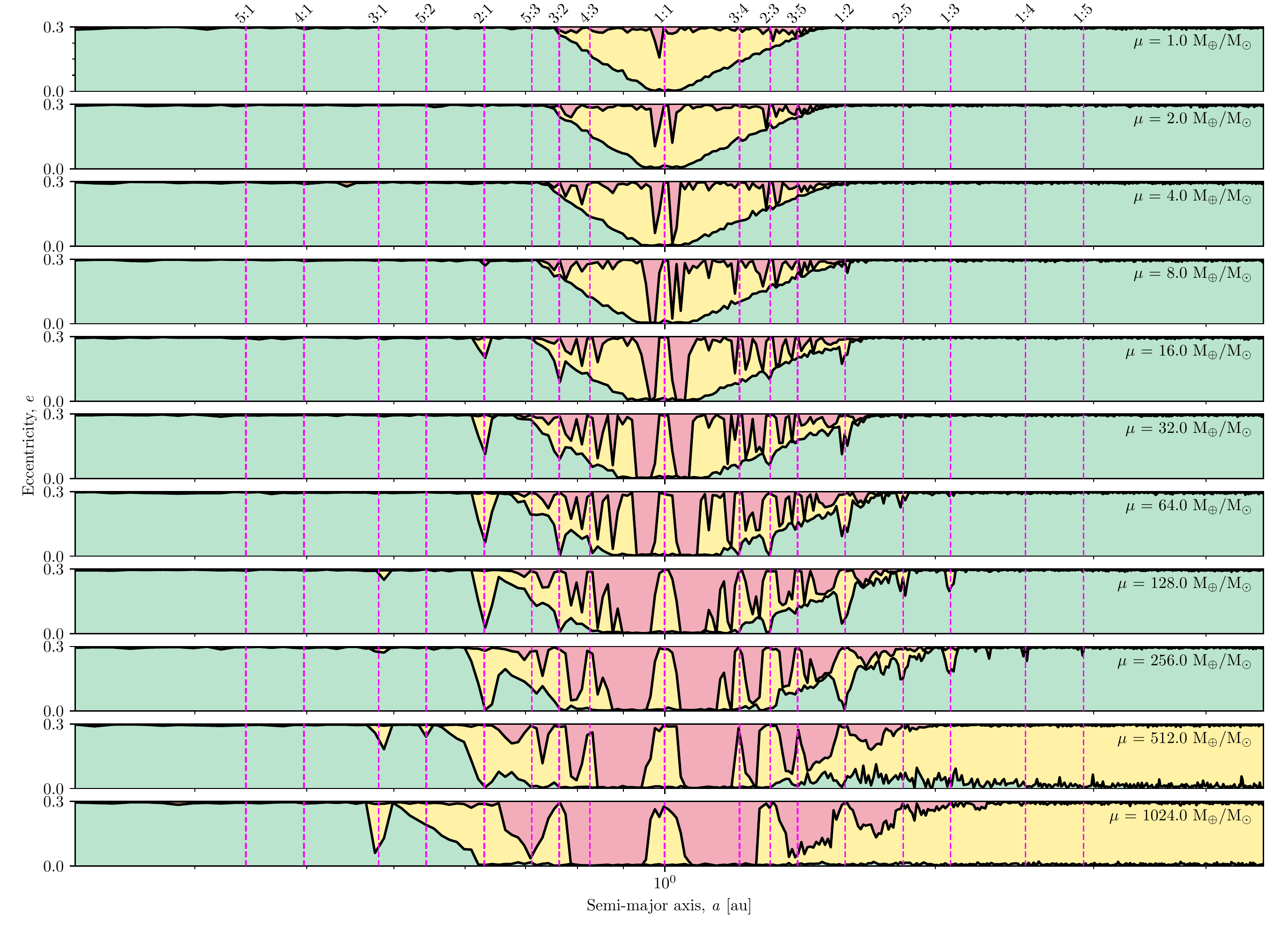}
    	\caption{All stability signatures for the $\mu$ values tested in simulation Set 1}\label{fig:all_signals}
	\end{subfigure}
	
    \begin{subfigure}{\textwidth}
		\centering
    	\includegraphics[width=\linewidth]{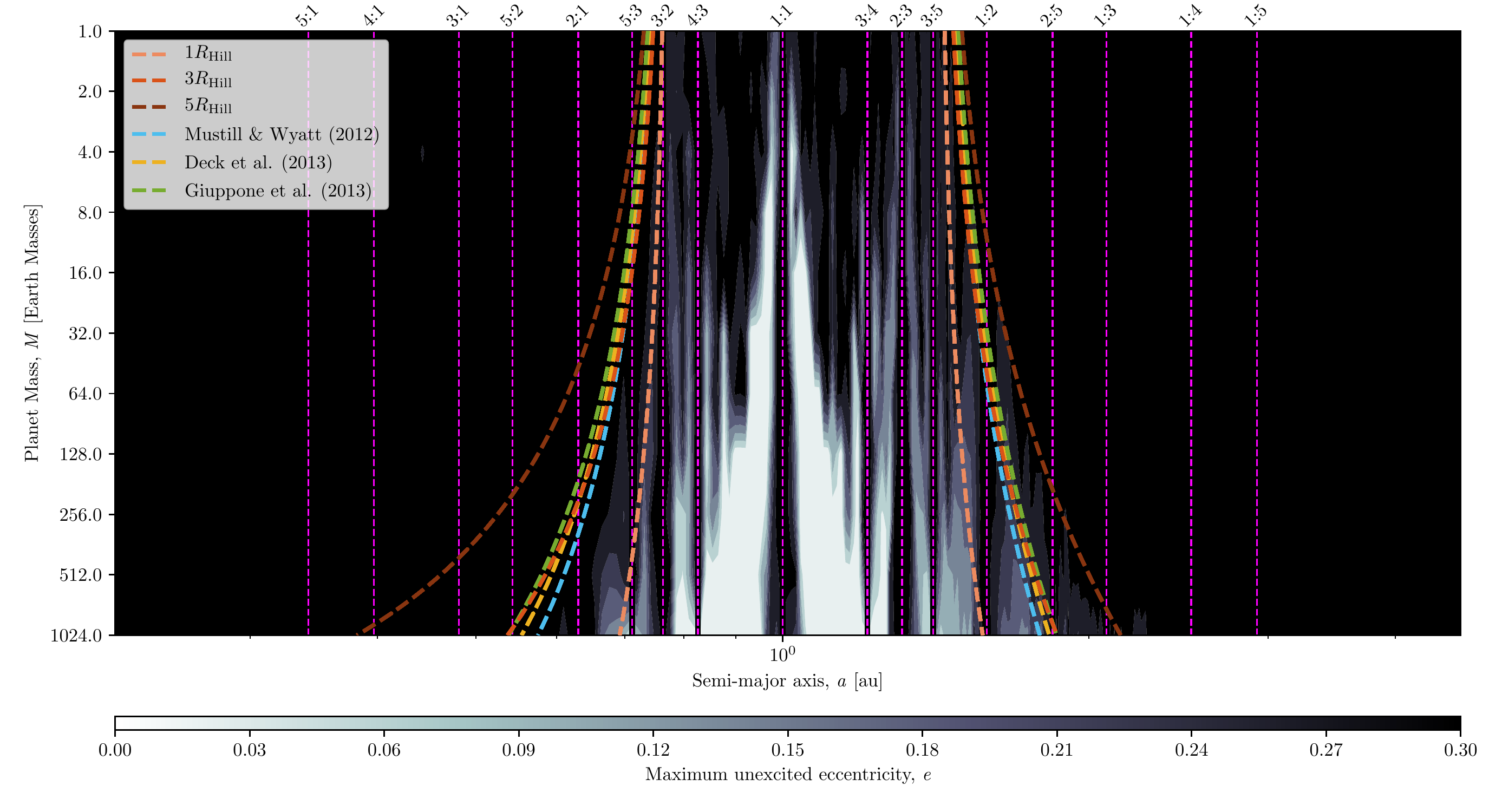}
    	\caption{Optimistic look-up map for interpolation}\label{fig:contour}
    \end{subfigure}
\end{figure*} 	
\begin{figure*}
    \ContinuedFloat
    \begin{subfigure}{\textwidth}
		\centering
    	\includegraphics[width=\linewidth]{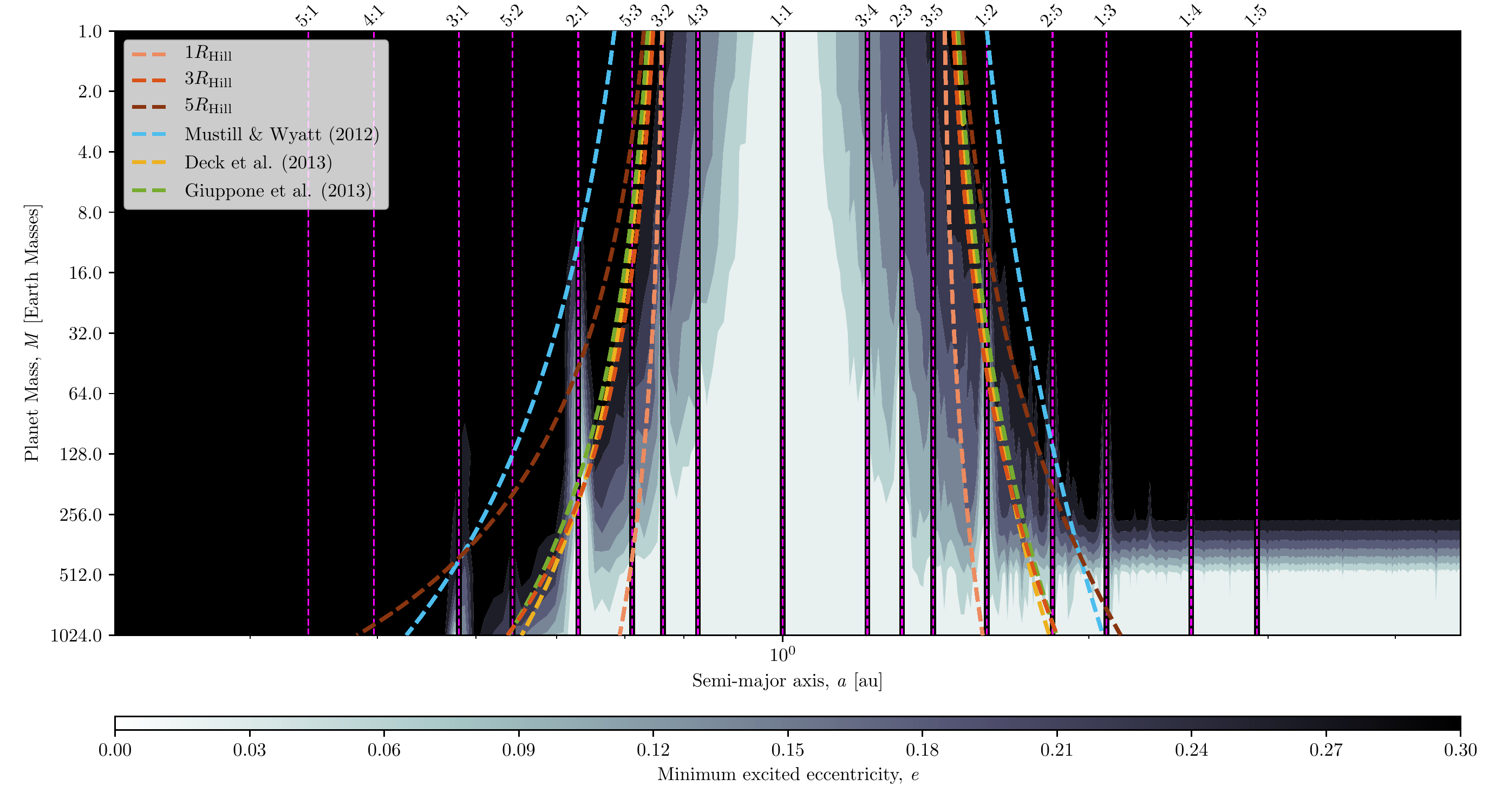}
    	\caption{Conservative look-up map for interpolation}\label{fig:contour_r}
    \end{subfigure}
    \caption{The (a) stability signatures, (b) optimistic look-up map, and (c) conservative look-up map, of the eleven planetary masses simulated between $\mu={1 \textrm{M}_{\oplus}}/\textrm{M}_\odot$ and $\mu={1024 \textrm{M}_{\oplus}}/\textrm{M}_\odot$. In these simulations, the systems considered consist of TPs on initially low eccentricity ($e<0.3$) orbits that are co-planar with the massive planet that perturbs them. Fig~\ref{fig:all_signals} shows the stability signature for each planetary mass simulated. The red region is the unstable region, the green is the stable region, and the yellow is in between where demonstrably stable and unstable TPs have been shown to exist. Figures~\ref{fig:contour} and \ref{fig:contour_r} combine all the stability signatures into contour maps, allowing for stability signatures at interim masses to be obtained by interpolating logarithmically between those we simulated.}\label{fig:mass_inf}
\end{figure*}

We emphasise that for low mass planets the optimistic stability signatures are difficult to extract accurately without making significant assumptions or changes to the extraction methodology. This was demonstrated for $\mu={8 \textrm{M}_{\oplus}}/\textrm{M}_\odot$ in Figure~\ref{fig:m8_extract}, and can be seen for other masses $\mu\leq{8 \textrm{M}_{\oplus}}/\textrm{M}_\odot$ in Figure~\ref{fig:all_signals}. Despite this, the conservative signature for such low mass scenarios can still be seen to be demonstrably stable for all but the regions located nearest to the planet, but this should be kept in mind for any systems with $\mu\leq{8 \textrm{M}_{\oplus}}/\textrm{M}_\odot$. 

Figure~\ref{fig:all_signals} shows that for systems with $\mu>{8 \textrm{M}_{\oplus}}/\textrm{M}_\odot$, the optimistic stability signatures are much less noisy and hence more distinctly defined. The semi-major axes that correspond with stabilising mean motion resonances are also clearly visible (shown by the dashed magenta lines), as is the manner in which the widths of those resonances can vary with planetary mass. In contrast, Figure~\ref{fig:all_signals} shows that the conservative signals are much more distinctly defined even for low mass ratios, and so are useful for identifying stability.

Figures~\ref{fig:contour} and \ref{fig:contour_r} shows the look-up maps we developed by representing the eccentricity values of the stability signature for each $\mu$ as shaded areas. The dark regions represent strongly stable regions, specifically where the maximum eccentricity of an unexcited TP is $0.3$ (the maximum value we used in our simulations). Conversely, the lighter regions represent the unstable regions, where the maximum eccentricity of an unexcited TP is zero or near zero, meaning that region in ($a$,$e$) space has been completely cleared of TPs. The stabilising resonances are again evident in Figure~\ref{fig:contour} (the vertical dark streaks embedded in the light white cloud, highlighted by the dashed magenta lines), as is the strong influence that the mass ratio has on the reach of the unstable region. We see similar features in Figure~\ref{fig:contour_r}, but in this case the mean motion resonances can be seen to have a destabilising effect on the TPs, specifically the 2:1 and 1:2 MMRs. In general, the various analytic definitions for the onset of chaos bounds or straddles the unstable regions we present to some extent in both the optimistic and conservative cases, suggesting there is use to both the optimistic and conservative definitions depending on the desired application. It also highlights that a more robust definition of excitation utilising the derivations in these works may yield better, more precisely defined signatures. Generally, these analytic criteria derive the boundary of the onset of chaos to be related to the relative difference in semi-major axis between the two bodies. As such, monitoring this throughout each simulation may yield a far more robust method by which to obtain the stability signatures. The results of each simulation from which the stability signature was extracted can be found in Figures~\ref{fig:mass_effect}. 

In addition to the details that are evident in Figures~\ref{fig:contour} and \ref{fig:contour_r}, the look-up map can also be used to assess dynamical stability in a system, with specific emphasis on the conservative stability contour map. We can interpolate between the masses we simulated to obtain the signature for a planet of any mass (within the maximum and minimum mass ratios of $\mu=1-{1024 \textrm{M}_{\oplus}}/\textrm{M}_\odot$). We look at how to do this to achieve our two goals -- assessing dynamical stability in multiple planet systems, and how to identify systems where an exo-Earth may exist in the HZ of systems -- in section~\ref{sec:appls}.

\subsection{Effect of eccentricity}
\label{subsec:ecc}
\begin{figure*}
	\begin{subfigure}{\textwidth}
		\centering
    	\includegraphics[width=\linewidth]{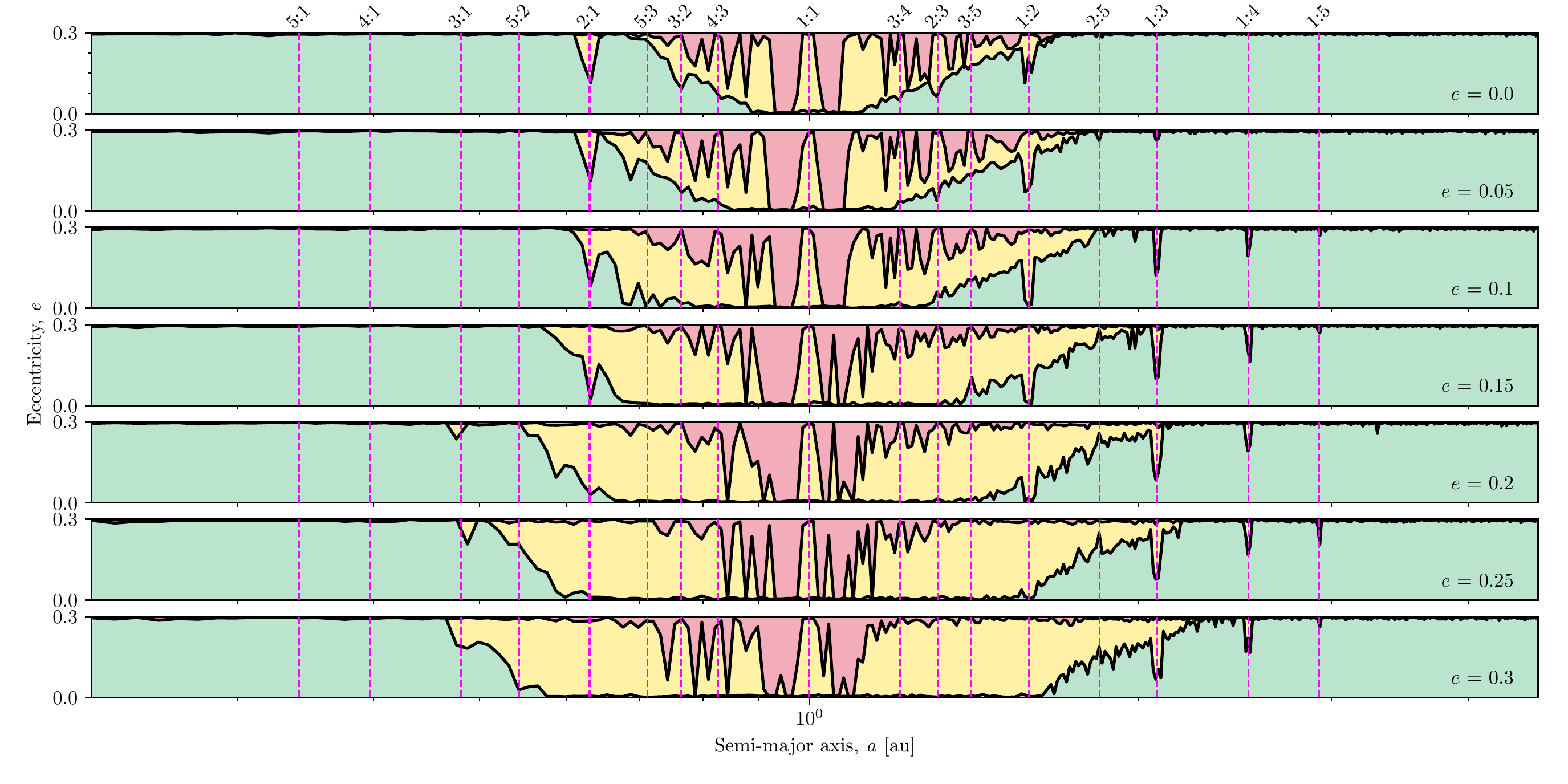}
    	\caption{All stability signatures for the $e$ values tested in simulation Set 2}\label{fig:e_signals}
	\end{subfigure}
	
    \begin{subfigure}{\textwidth}
		\centering
    	\includegraphics[width=\linewidth]{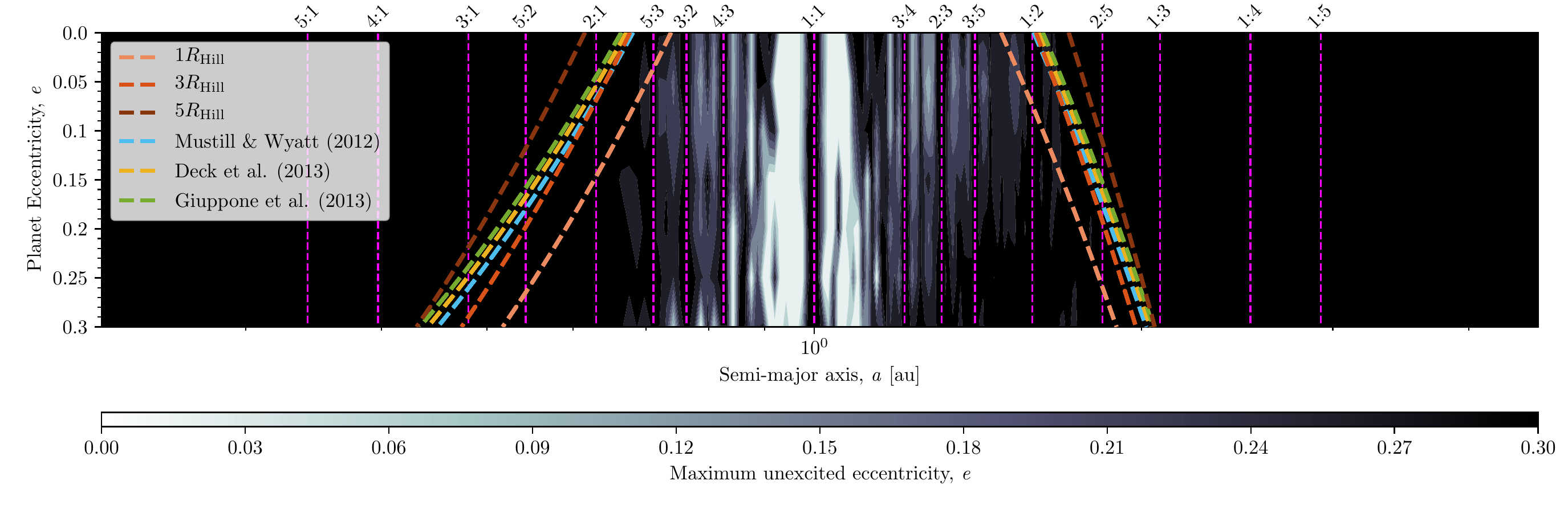}
    	\caption{Optimistic look-up map for interpolation}\label{fig:e_contour}
    \end{subfigure}
	\begin{subfigure}{\textwidth}
		\centering
    	\includegraphics[width=\linewidth]{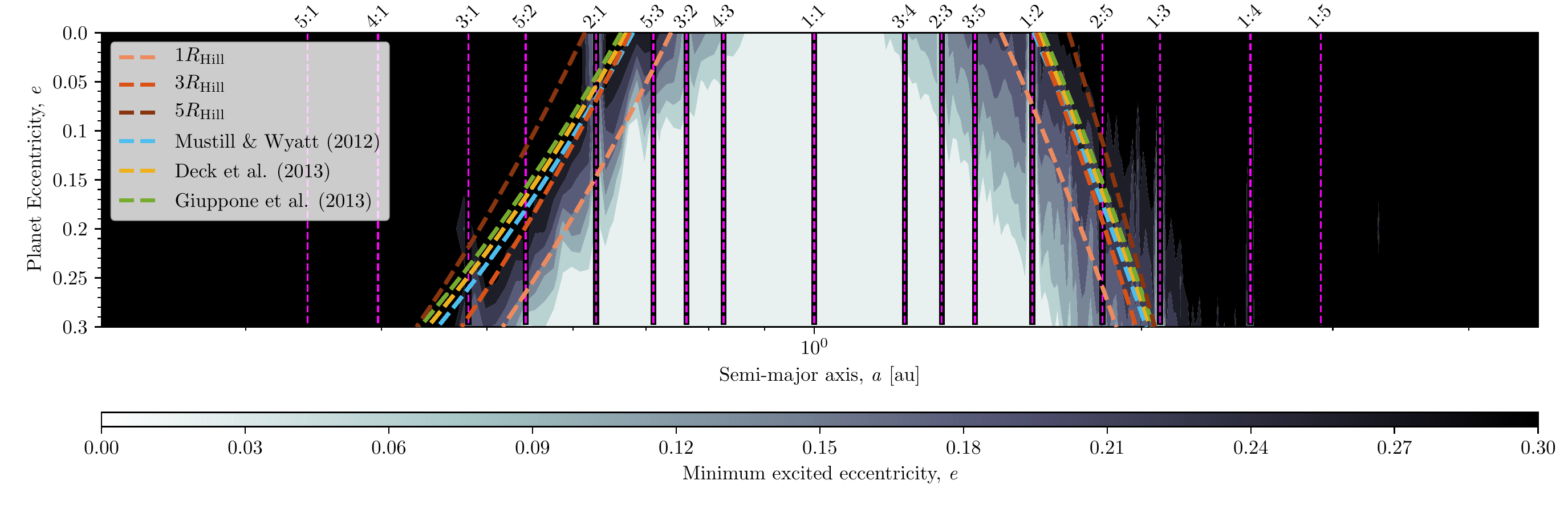}
    	\caption{Conservative look-up map for interpolation}\label{fig:e_contour_r}
    \end{subfigure}	
    \caption{The (a) stability signatures, (b) optimistic look-up map and (c) conservative look-up map for the seven planetary eccentricities simulated between $e=0$ and $e=0.3$. In these simulations, the systems are co-planar and the planet has mass ratio $\mu={32 \textrm{M}_{\oplus}}/\textrm{M}_\odot$. Fig~\ref{fig:e_signals} shows the stability signature for each planetary mass simulated. The red region is the unstable region, the green is the stable region, and the yellow is inbetween where demonstrably stable and unstable TPs have been shown to exist. Figures~\ref{fig:e_contour} and \ref{fig:e_contour_r} combine all the stability signatures into contour maps, allowing for stability signatures at interim eccentricities to be obtained by interpolating between those we simulated.}\label{fig:ecc_inf}
\end{figure*}

Whilst we focus primarily on co-planar, circular orbits, our method can be further developed and refined to extend to eccentric orbits. In this work, the eccentricity look-up map we generate is limited to a single mass ratio ($\mu = 32 \textrm{M}_{\oplus}/\textrm{M}_\odot$) and is not used like the ones presented in Figure~\ref{fig:mass_inf}. However, it does allow us to explore the influence of planetary eccentricity on the stability of a system. Figure~\ref{fig:ecc_inf} shows the optimistic and conservative stability signatures, and the look-up maps for a mass ratio of $\mu = 32 \textrm{M}_{\oplus}/\textrm{M}_\odot$. The $x$-axis shows the semi-major axis range simulated to obtain the stability signatures, i.e. between orbits with periods in $10:1$ and $1:10$ commensurability with the planet (which for $a_\mathrm{pl} = 1~\mathrm{au}$ is given by $0.215~\mathrm{au}\lesssim a\lesssim4.642~\mathrm{au}$), while the $y$-axis shows the eccentricity values of the stability signatures in Figure~\ref{fig:e_signals}, and indicates the eccentricity of the planet in each simulation in Figures~\ref{fig:e_contour} and \ref{fig:e_contour_r}. We only explore between $e=0$ and $e=0.3$, given that existing literature shows that multiple planet systems (especially where one is a potentially habitable terrestrial planet) are uncommon for high eccentricity orbits ($e > 0.4$). This has been explained as being most likely attributable to planet-planet scattering during planetary formation and evolution \citep{Matsumura2016,Carrera2016,Agnew2017,Zinzi2017}.

The stability signatures resulting from these simulations prove more challenging to extract since the TPs become far more excited. This can be seen by plotting the excitation of the TPs in ($a$,$e$) space in Figure~\ref{fig:mass_effect} and Figure~\ref{fig:ecc_effect} for the circular and non-circular cases respectively. While we do see the theoretical work bounding the unstable, chaotic region in Figure~\ref{fig:e_contour_r}, there is  a noticeable deviation with the optimistic signature (as evident in Figure~\ref{fig:e_contour}). This highlights two important notions: 1) as mentioned earlier, the case for investigating the effect of eccentricity means the system is no longer a circular restricted three-body problem. As such, the Jacobi integral is not conserved and so we see significantly different dynamics than in the circular case when investigating the effect of mass, and 2) the method by which we determine excitation is not well suited when investigating eccentric orbits. Since we see qualitative agreement in Figures~\ref{fig:contour}, \ref{fig:contour_r} and \ref{fig:e_contour_r} with different analytic derivations for the onset of chaos \citep[e.g.][]{Mustill2012, Deck2013,Giuppone2013}, a modification of the excitation criteria that utilises some of the research presented in these works should be incorporated in future work that seeks to include the eccentricity parameter space. This may yield criteria that are more suitable for the eccentric cases. %These analytic criteria define the boundary of the onset of chaos to be related to the relative semi-major axis between the two bodies, and so such a criteria may be far more suitable for the eccentric cases.

Figure~\ref{fig:ecc_inf} demonstrates the destabilising influence planetary eccentricity has on the dynamical stability of nearby objects -- a result that matches what has been found in previous studies of proposed multiple planet systems \mbox{\citep{Carrera2016,Agnew2017}}. Figure~\ref{fig:ecc_inf} shows how rapidly the regions nearby a planet are destabilised as the planet moves from a circular to an eccentric orbit. The semi-major axes that correspond to the locations of potential stabilising mean motion resonances are also far less pronounced, as seen by the less distinct dark vertical streaks in Figure~\ref{fig:e_contour}. Even more impressive is the range of destabilisation in semi-major axis in Figure~\ref{fig:e_contour_r}, reaching out in both directions to the locations of the $3:1$ inner resonance and the $1:4$ outer resonance with the planet's orbit in the highest eccentricity case we examined ($e_\mathrm{pl}=0.3$). In contrast, the unstable regions in the circular case only reached to the $5:3$ inner, and just beyond the $1:2$ outer respectively, re-enforcing the significance of even moderate orbital eccentricities on multiple system stability \citep{Zinzi2017}.

\subsection{Effect of inclination}
\label{subsec:inc}
\begin{figure*}
    \begin{subfigure}{\textwidth}
		\centering
    	\includegraphics[width=\linewidth]{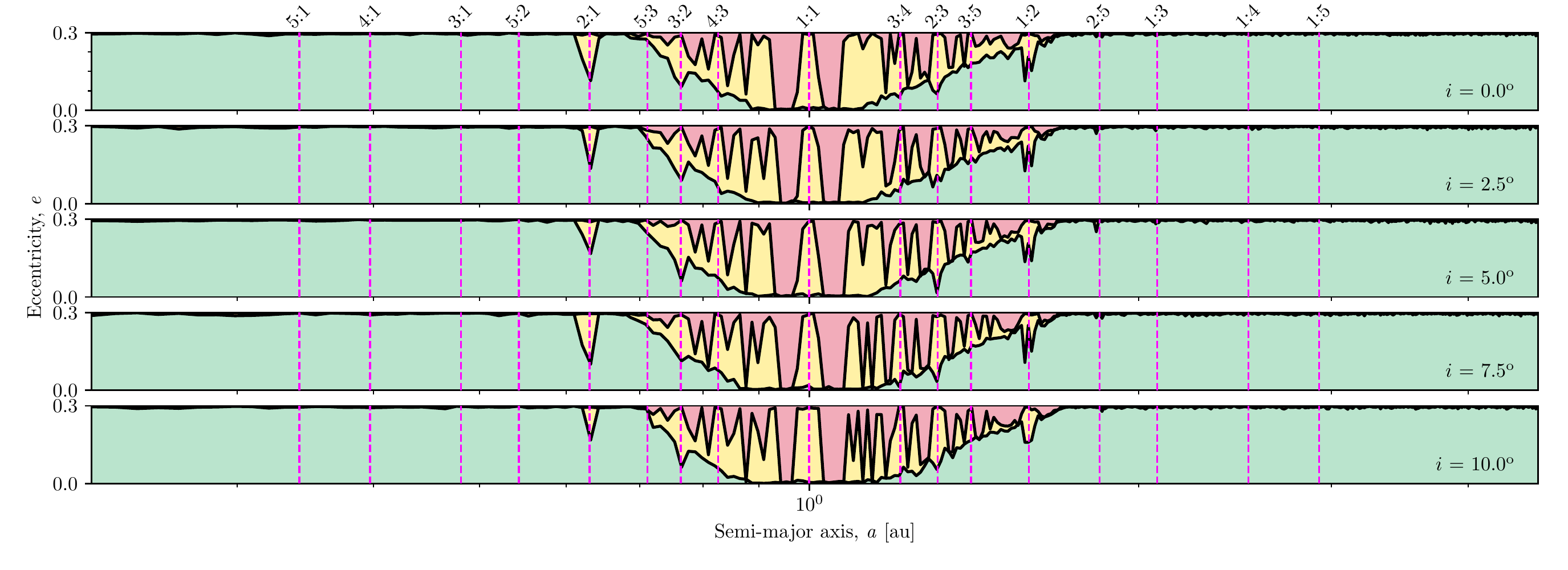}
    	\caption{All stability signatures for the $i$ values tested in simulation Set 3}\label{fig:i_signals}
	\end{subfigure}

    \begin{subfigure}{\textwidth}
		\centering
    	\includegraphics[width=\linewidth]{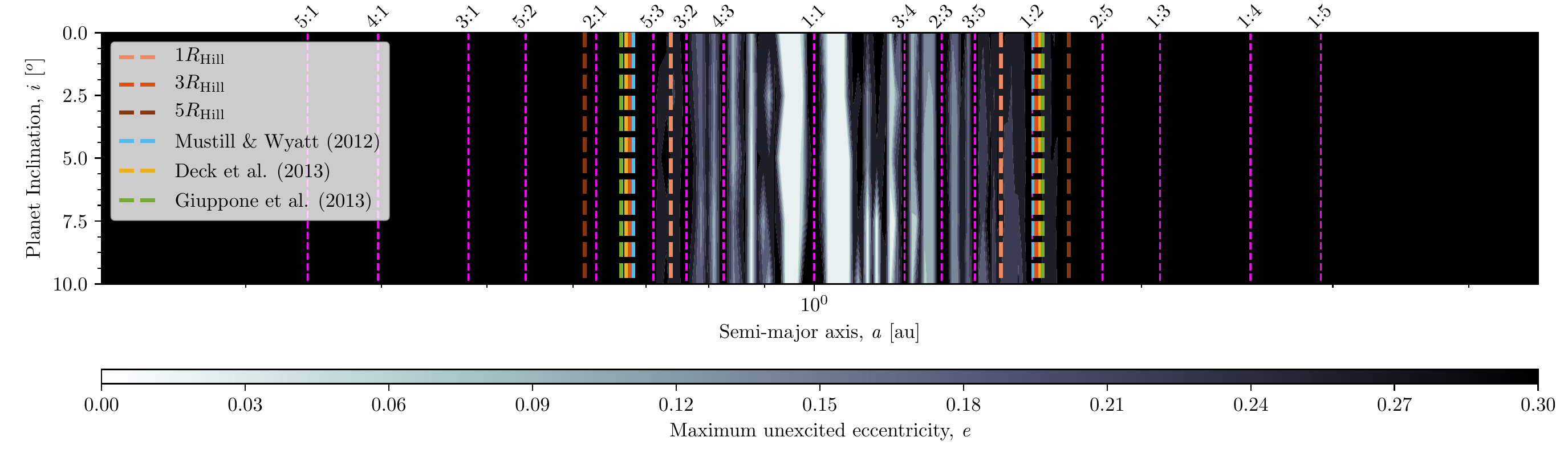}
    	\caption{Optimistic look-up map for interpolation}\label{fig:i_contour}
    \end{subfigure}
    
	\begin{subfigure}{\textwidth}
		\centering
    	\includegraphics[width=\linewidth]{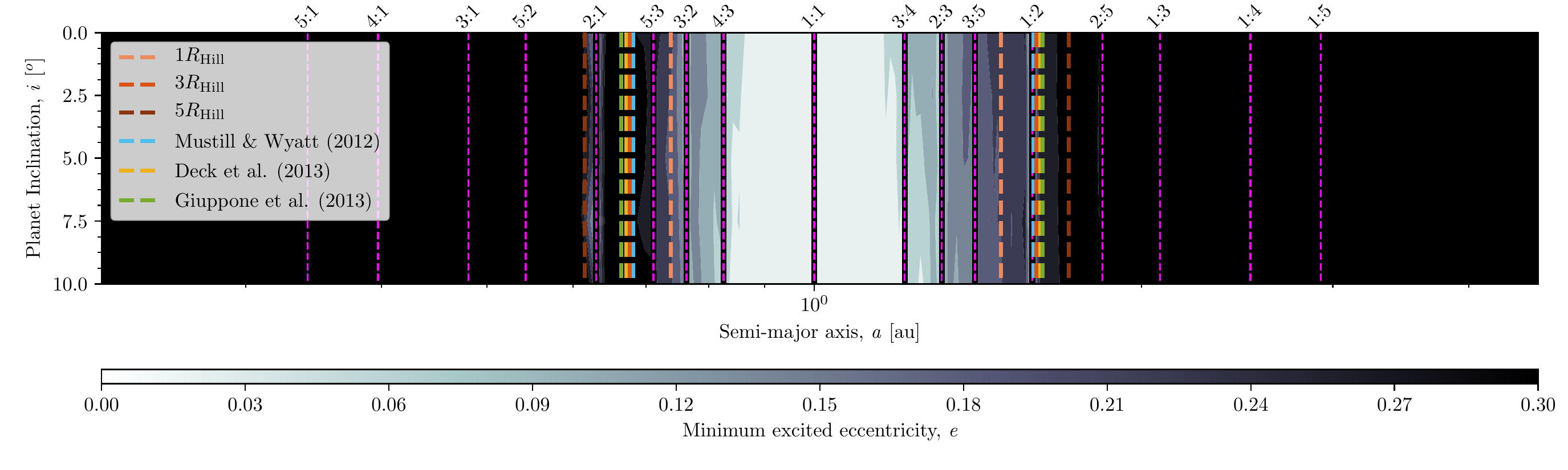}
    	\caption{Conservative look-up map for interpolation}\label{fig:i_contour_r}
    \end{subfigure}
    \caption{The (a) stability signatures, (b) optimistic look-up map and (c) conservative look-up map for the various planetary inclinations simulated between $i=0\degree$ and $i=10\degree$. In these simulations, the planet is on a circular orbit and has mass ratio $\mu={32 \textrm{M}_{\oplus}}/\textrm{M}_\odot$. Fig~\ref{fig:i_signals} shows the various stability signatures for each planetary inclination simulated. The red region is the unstable region, the green is the stable region, and the yellow is inbetween wheredemonstrably stable and unstable TPs have been shown to exist. Fig~\ref{fig:i_contour} and \ref{fig:i_contour_r} combines the stability signatures into contour maps, allowing for stability signatures at interim inclinations to be obtained by interpolating between those we simulated.}\label{fig:inc_inf}
\end{figure*}

Similar to our investigation into the effect of eccentricity on the stability signatures of a planet, here we explore the influence of planetary inclination on the stability of a system. Figure~\ref{fig:inc_inf} shows the various stability signatures and the look-up maps for a mass ratio of $\mu = 32 \textrm{M}_{\oplus}/\textrm{M}_\odot$. The $x$-axis shows the semi-major axes that we simulated to obtain the stability signatures, while the $y$-axis shows the eccentricity values of the stability signatures in Figure~\ref{fig:i_signals}, and indicates the inclination of the planet in each simulation in Figures~\ref{fig:i_contour} and \ref{fig:i_contour_r}.

What is immediately apparent from both Figures~\ref{fig:i_signals} and \ref{fig:i_contour} is that shallow mutual inclinations (i.e. $i\leq10\degree$) have very little influence on the stability signatures of a system. Figure~\ref{fig:i_signals} shows that the stability signatures vary very little outside of the inherent stochastic variations expected using a randomly distributed swarm of TPs. Figure~\ref{fig:i_contour} shows that the stabilising influence of mean motion resonances remains more or less consistent across the range of inclinations explored, as is evident by the near uniform vertical dark streaks. Similarly, the destabilising effect of the 2:1 and 1:2 mean motion resonances are evident across all inclinations, with the conservative signature in Figure~\ref{fig:i_contour_r} showing depletion in regions that align with the locations of the 2:1 and 1:2 MMRs.

While the conditions here are those for the circular restricted three-body problem, it is important to note that the dynamics will change still further when the third body is treated as a massive, rather than massless object -- in other words when we move from considering the restricted three-body problem to the unrestricted three-body problem. The mutual gravitational interactions that would exist between massive bodies, rather than a massive body and massless TPs, is expected to yield more complicated dynamical behaviour. In such a scenario, inclination would have a much more significant influence on dynamical stability. However, previous studies have shown that multiple planet systems are likely to exist on shallow, near co-planar orbits \citep{Lissauer2011,Lissauer2011a,Fang2012,Figueira2012,Fabrycky2014}.

\section{Applications}
\label{sec:appls}
The first batch of \textit{TESS} science observations has already resulted in the detection of two planets \citep{Huang2018,Vanderspek2018}, and this is expected to grow to several thousand throughout the lifetime of the survey \mbox{\citep{Ricker2014,Sullivan2015,Barclay2018}}. Here, we demonstrate how our approach can be applied for our two proposed use cases for newly discovered \textit{TESS} systems. 

The first use case is to conservatively assess the dynamical stability of multiple planet systems. The approach we take is to assess the stability of each planet separately to provide insight into the multiple planet system as a whole. We detail this in section~\ref{app_stab}. The assessments presented here are conducted using the best-fit orbital parameters inferred from observations to test the robustness of our approach. We follow this with a demonstration of how to assess stability across the region covered by the uncertainties of the orbital parameters of planets, and so how our method can provide a means to assess the stability of a system with the currently inferred planetary parameters, and if more observations are needed to better constrain the true nature of the system \citep[e.g.][]{Anglada-escud2010,Anglada-Escude2010,Wittenmyer2013,Horner2014}. As our look-up map in Figures~\ref{fig:contour} and \ref{fig:contour_r} are only for circular orbits, and does not take into account secular interactions which can occur over longer timescales, this is a conservative assessment that can be used as a quick, first order check of a system, enabling a rapid assessment of systems that are likely to be dynamically unstable with their current nominal best-fit orbits. 

The secondary use case allows for the rapid identification of single planet (and certain multiple planet) systems where additional planets can maintain stable orbits within the HZ (with potential Earth-mass planets being of particular interest here). In this way, the expected return of several thousand newly discovered systems by \textit{TESS} can be quickly assessed to identify those which could contain as yet undetected exo-Earths. We detail how we achieve this in section~\ref{app_HZ}.

\subsection{Multiple Planet Stability}\label{app_stab}
\begin{figure}
% 	\begin{subfigure}{\linewidth}
	\centering
    \includegraphics[width=\linewidth]{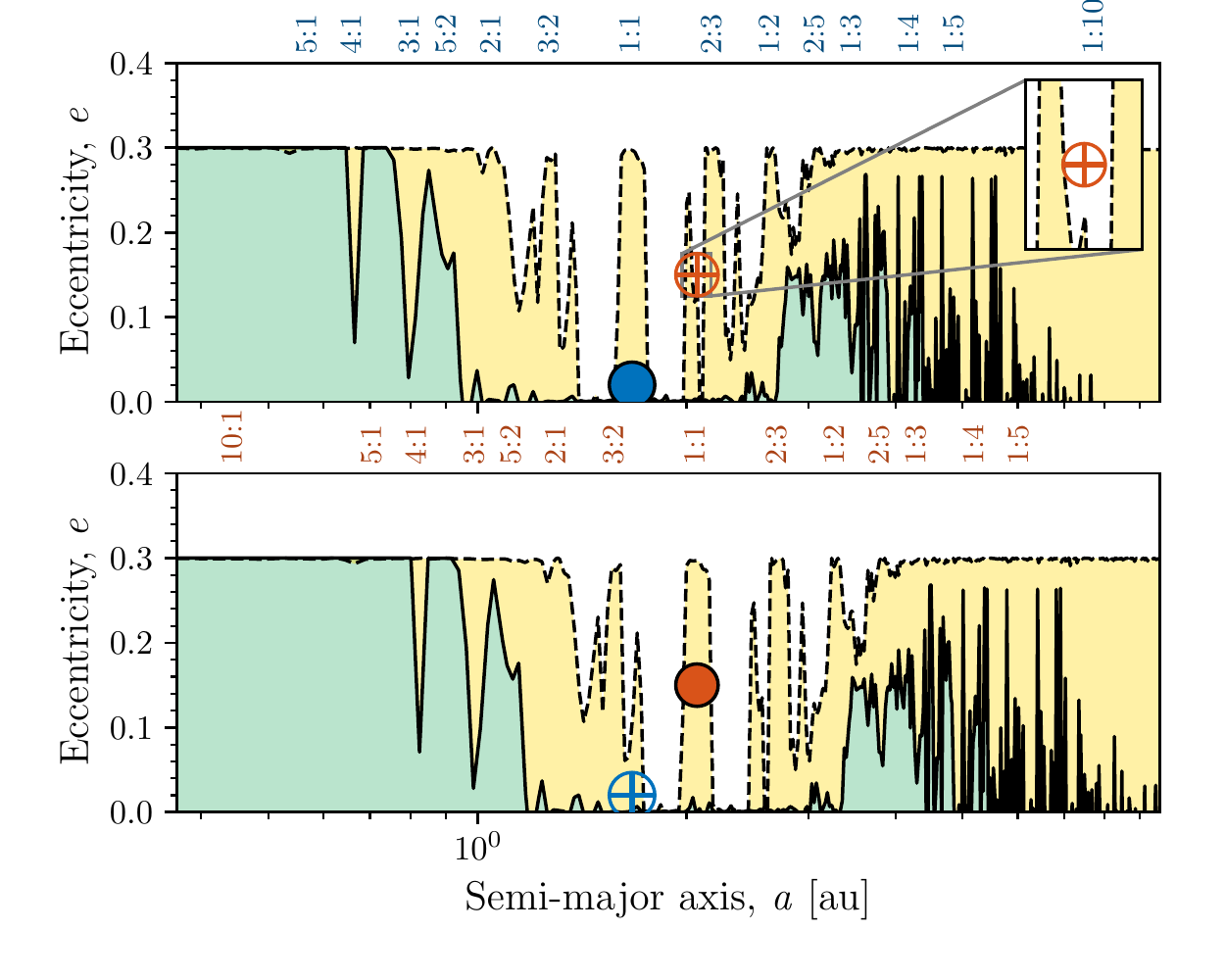}
    % \caption{Individual stability signatures}\label{fig:insta_pred_ind}
% 	\end{subfigure}

% 	\begin{subfigure}{\linewidth}
% 		\centering
%     	\includegraphics[width=\linewidth]{images/predictions/hd_5319_combined.pdf} 
%     	\caption{Combined stability signatures}\label{fig:insta_pred_comb}
%  	\end{subfigure}
	\caption{A comparison between the predictions made using our method, and a previously run numerical simulation conducted by \protect\cite{Agnew2018b} for HD~5319. For each planet we overlay its stability signature and assess the location of the other planet with respect to the first. In (a) the red planet is located above the optimistic stability signature, suggesting instability. In (b) the blue planet is located within the potentially stable region. Despite that, as one of the planets is considered unstable, it can be suggested these particular system parameters are unstable.}
	\label{fig:instability_predictions}	
\end{figure}
% \begin{figure}
% % 	\begin{subfigure}{\linewidth}
%     \centering
%     \includegraphics[width=\linewidth]{images/predictions/stab_hip_65407.pdf}
% %     	\caption{Individual stability signatures}
% % 	\end{subfigure}

% % 	\begin{subfigure}{\linewidth}
% % 		\centering
% %     	\includegraphics[width=\linewidth]{images/predictions/hd_141399_combined.pdf}
% %     	\caption{Combined stability signatures}
% % 	\end{subfigure}
% 	\caption{{A comparison between the predictions made using our method, and a previously run numerical simulation conducted by \protect\cite{Agnew2018b} for HIP~65407. In (a) the coloured circles represent the planets and the coloured curves represent each planets stability signature. In (b) outlined circles and crosses are represent the planets, and the green curve represents the combined stability signature. Outlined circles and crosses are used so as to more easily compare where the planet is located with respect to the stability signature. The shaded green region represents the HZ.}}
%     \label{fig:stability_predictions}
%\end{figure}

\begin{figure}
% 	\begin{subfigure}{\linewidth}
    \centering
    \includegraphics[width=\linewidth]{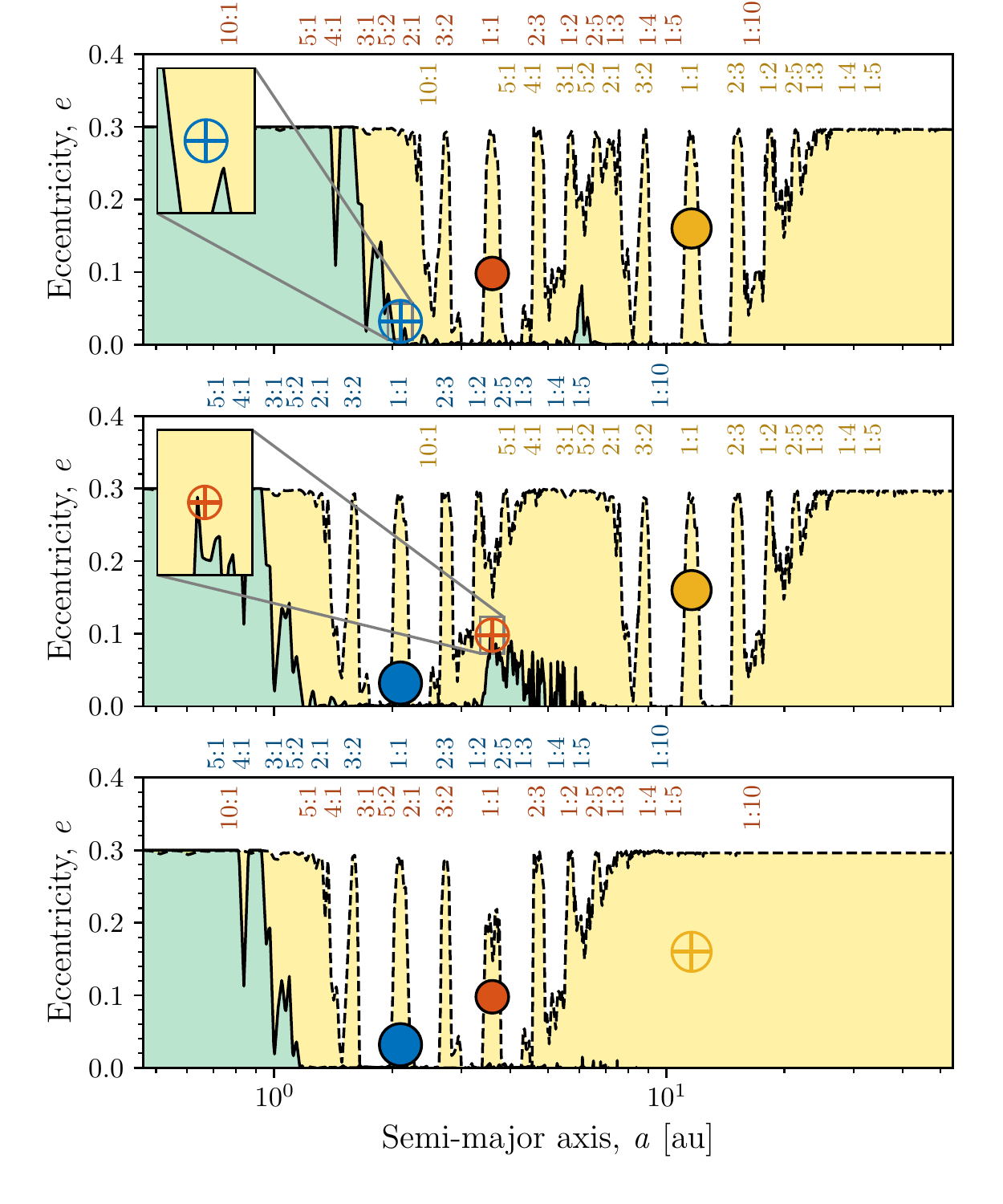}
%     	\caption{Individual stability signatures}
% 	\end{subfigure}

% 	\begin{subfigure}{\linewidth}
% 		\centering
%     	\includegraphics[width=\linewidth]{images/predictions/hd_141399_combined.pdf}
%     	\caption{Combined stability signatures}
% 	\end{subfigure}
	\caption{A comparison between the predictions made using our method, and a previously run numerical simulation conducted by \protect\cite{Agnew2018b} for 47~UMa. In each panel, one of the planets stability (the circle and cross marker) is assessed based on where it is located relative to the combined stability signature of the other two planets (the solid markers). In this system, all the planets potentially stable, so it can be suggested that this particular system may be stable.}
    \label{fig:stability_predictions_mult}
\end{figure}

To assess the stability of multiple planet systems, the stability of each planet needs to be assessed  separately. Thus, for any planet, $P_1$, we must investigate the gravitational influence that the other planets, $P_2$, $P_3$, $P_4$ ..., $P_n$ have on $P_1$. The same assessment must be conducted for each planet in the system, investigating the gravitational influence that the other planets $P_1$, $P_3$, $P_4$, ..., $P_n$ have on $P_2$; the gravitational influence that the other planets $P_1$, $P_2$, $P_4$, ..., $P_n$ have on $P_3$; and so on.

Let us consider a three planet system. Starting with $P_1$, we first obtain the stability signature of the other planets by combining their masses with $P_1$ (i.e. with respective masses   $M_2'=M_1+M_2$ and $M_3'=M_1+M_3$), interpolating within our look-up maps, and translating the signatures to each planet's semi-major axis $a_2$ and $a_3$ respectively. We combine these signatures to get an optimistic and a conservative signature by taking the minimum value of the $P_2$ and $P_3$ stability signatures at each semi-major axis location. We then plot these combined stability signatures, as well as where $P_1$ is located, in ($a$, $e$) parameter space. We can then infer if $P_1$ would be stable, unstable, or potentially stable based on where it lies within the combined stability signature. The inference of stability is determined as outlined in section~\ref{sec:method}: below the conservative line corresponds with the unexcited, stable regions around a planet; above the optimistic line corresponds with the excited regions around a planet; and in between corresponding with potentially stable regions. This process is then repeated for $P_2$ and $P_3$. Once all planets in the multiple system have been assessed, if any planet is located above the optimistic stability signature, the system is considered potentially unstable. Conversely, if all of them fall below the conservative stability signature, the system can be considered stable. A worked example of this process for HD~5319 can be found in Appendix~\ref{sec:appendix_examples}.

We test the robustness our stability signature predictions with the dynamical stability of systems found through numerical simulations. \cite{Agnew2018b} conducted such a study, simulating all multiple planet systems with a gas giant within the then known exoplanet population for $10^8$ years in order to assess their dynamical stability. Figure~\ref{fig:instability_predictions} shows the HD~5319 system, which \cite{Agnew2018b} found to be dynamically unstable. We can see that the red planet is located above the optimistic stability signature of the blue planet, meaning it resides in the excited, unstable regions within the system, suggesting the system overall is unstable. In Figure~\ref{fig:stability_predictions_mult} we show a similar assessment of the 47~UMa system which was found to be potentially dynamically stable by \cite{Agnew2018b}. We follow the method outlined earlier for assessing the stability of a three body system by combining stability signatures, and we can see that all three planets are located within the potentially stable region.

We can compare our stability signature predictions with all the multiple planet systems that \cite{Agnew2018b} tested numerically. As they sought to identify systems that may host hidden exo-Earths in the HZ, they had additional criteria relating to the habitability of each system that ultimately yielded a selection of 51 multiple planet systems from the then known exoplanet population. As our stability signatures only cover mass ratios $\mu \leq 1024 \textrm{M}_{\oplus}/\textrm{M}_\odot$, and $e\leq0.3$, this places a limitation on which systems we can test our predictions against. Of the 51 systems \cite{Agnew2018b} tested numerically, 25 systems fall within the mass ratio and eccentricity ranges we used here, and so we we can directly compare our stability signature predictions with these 25 systems. Our comparison is summarised in Table~\ref{tab:stability_pred}.

\begin{table}
\caption{Multiple exoplanet systems stability signature predictions compared with the stability of the system determined from detailed numerical simulations of \protect\cite{Agnew2018b}. This table is intended to demonstrate the agreement between our approach and numerical simulations, but should not be seen as a definitive assessment of those system's stability.}
	\label{tab:stability_pred}	
    \centering
    \begin{tabular}{l c c c}
        \toprule                     
        System		& Numerical & Signature	& Agreement?\\
        			& Result 	& Prediction& \\
        \midrule
        24 Sex		&	Unstable	&	Unstable	&	\cmark	\\
		47 UMa		&	Stable	&	Potentially Stable	&	-	\\
		BD-08 2823	&	Stable	&	Stable	&	\cmark	\\
		HD 10180	&	Stable	&	Potentially Stable	&	-	\\
		HD 108874	&	Stable	&	Potentially Stable	&	-	\\
		HD 113538	&	Stable	&	Potentially Stable	&	-	\\
		HD 134987	&	Stable	&	Potentially Stable	&	-	\\
		HD 141399	&	Stable	&	Potentially Stable	&	-	\\
		HD 142		&	Stable	&	Potentially Stable	&	-	\\
		HD 159868	&	Stable	&	Potentially Stable	&	-	\\
		HD 1605		&	Stable	&	Potentially Stable	&	-	\\
		HD 160691	&	Unstable	&	Potentially Stable	&	-	\\
		HD 187123	&	Stable	&	Potentially Stable	&	-	\\
		HD 200964	&	Unstable	&	Unstable	&	\cmark	\\
		HD 219134	&	Stable	&	Stable	&	\cmark	\\
		HD 33844	&	Unstable	&	Unstable	&	\cmark	\\
		HD 47186	&	Stable	&	Stable	&	\cmark	\\
		HD 4732		&	Stable	&	Potentially Stable	&	-	\\
		HD 5319		&	Unstable	&	Unstable	&	\cmark	\\
		HD 60532	&	Stable	&	Potentially Stable	&	-	\\
		HD 9446		&	Stable	&	Potentially Stable	&	-	\\
		HIP 65407	&	Stable	&	Stable	&	\cmark	\\
		HIP 67851	&	Stable	&	Potentially Stable	&	-	\\
		TYC 1422-614-1	&	Stable	&	Potentially Stable	&	-	\\
		XO-2 S		&	Stable	&	Potentially Stable	&	-	\\
        \bottomrule
    \end{tabular}
\end{table}

We find that our stability signatures yield agreement on the dynamical stability of a system in $32\%$ (8/25) of the systems tested numerically by \cite{Agnew2018b}, no strong disagreement in any of the systems tested, and only one case of disagreement with a system found to be unstable numerically (HD~160691). This is a promising result, as our predictions are using stability signatures for circular orbits, and as highlighted in section~\ref{subsec:ecc}, higher eccentricity orbits will create less stable signatures. This means that the discrepancy in the numerical and signature predictions found for the HD~160691 (the only system \cite{Agnew2018b} found to be unstable that our method did not also predict to be unstable) may be reconciled, as the signature being used to predict stability would be more stable (due to it being circular) than what it should be in reality ($e>0$). It should be noted that our stability signatures do not take into account higher order secular interactions, and so stable predictions are inherently less conclusive because of these not being incorporated. There is also agreement in the potentially stable system predictions, although for multiple planet stability it is the unstable predictions that are more useful.

By demonstrating the robustness of our predictions, we can now present the use case for assessing the stability of a planetary system. We do this by carrying out Monte Carlo (MC) simulations for a system, randomly sampling within the accepted range of values. For each simulation, we can determine a stability metric by assigning a value of $0$ for an unstable system, $1$ for a stable system, and linearly interpolating the metric between $0$ and $1$ between the two signatures for the potentially stable systems (taking the minimum interpolated value of all planets in a given multiple system). Taking the mean of all the MC simulations for a system yields a measure of how stable the system is with the current planetary parameters. As an example, we consider HIP~65407, for which  planet HIP~65407~b has  $0.172\leq a_\mathrm{b}\leq0.182$~au and $0.396\leq M_\mathrm{b}\leq0.46$~$\mathrm{M}_\mathrm{J}$ and HIP~65407~c has $0.308\leq a_\mathrm{c}\leq0.324$~au and $0.73\leq M_\mathrm{c}\leq0.838$~$\mathrm{M}_\mathrm{J}$. We run 100,000 MC simulations and yield a stability metric of $92\%$, suggesting this system should not be prioritised when determining which systems require additional observations to better constrain the planetary parameters.

It must be re-iterated that this is a conservative prediction, and the implications of some dynamical interactions may be missed. Specifically, this assessment only compares stability between pairs of planets. For systems with more than two planets, this method will not include the effect of multiple-body interactions. Such interactions could potentially destabilise a planet-pair that would, on their own, be mutually stable. As a result, it seems likely that truly stable solutions would require planets to be somewhat more widely spaced in such systems than our two-planet stability maps might otherwise suggest \citep[e.g.][]{Pu2015,Pu2016, Tamayo2016}. Additionally, previous studies have demonstrated that the ratio of two planet's masses has very little influence on stability. Instead, it is the cumulative mass of the planets that impacts most upon their stability \citep{Petit2017,Hadden2018}. As such, our planetary predictions between large mass planets is optimistic as the application is more suited to identifying less massive, Earth-mass companions. The most appropriate use case for our approach in assessing planetary stability is in identifying unstable systems by assessing many permutations of planetary parameters across the constrained error bars, and using that assessment to inform observers as to whether to gain more data to better constrain them.

\subsection{Predicting HZ companions}\label{app_HZ}
Generally, detailed numerical simulations are the means by which to identify the stable and unstable regions within a planetary system \citep{Raymond2006a,Raymond2005,Kane2015}. Such simulations are computationally expensive and so other methods to more rapidly predict stability within a system would be particularly useful. Here, we demonstrate how we can utilise our approach to identify where additional planets can maintain stable orbits within the HZ -- with a specific focus on Earth-mass planets -- in lieu of computationally expensive simulations. It is essential to test the robustness of our approach, and we do so by comparing our predictions with the standard numerical approach. \cite{Agnew2017}, \cite{Agnew2018} and \cite{Agnew2018b} have performed such simulations to assess HZ stability (with both massless TPs and $1~\textrm{M}_{\oplus}$ bodies) for a large sample of single Jovian planet systems. Here we draw upon the results of their simulations to compare with our predictions. 

\begin{figure}
    \centering
    \includegraphics[width=\linewidth]{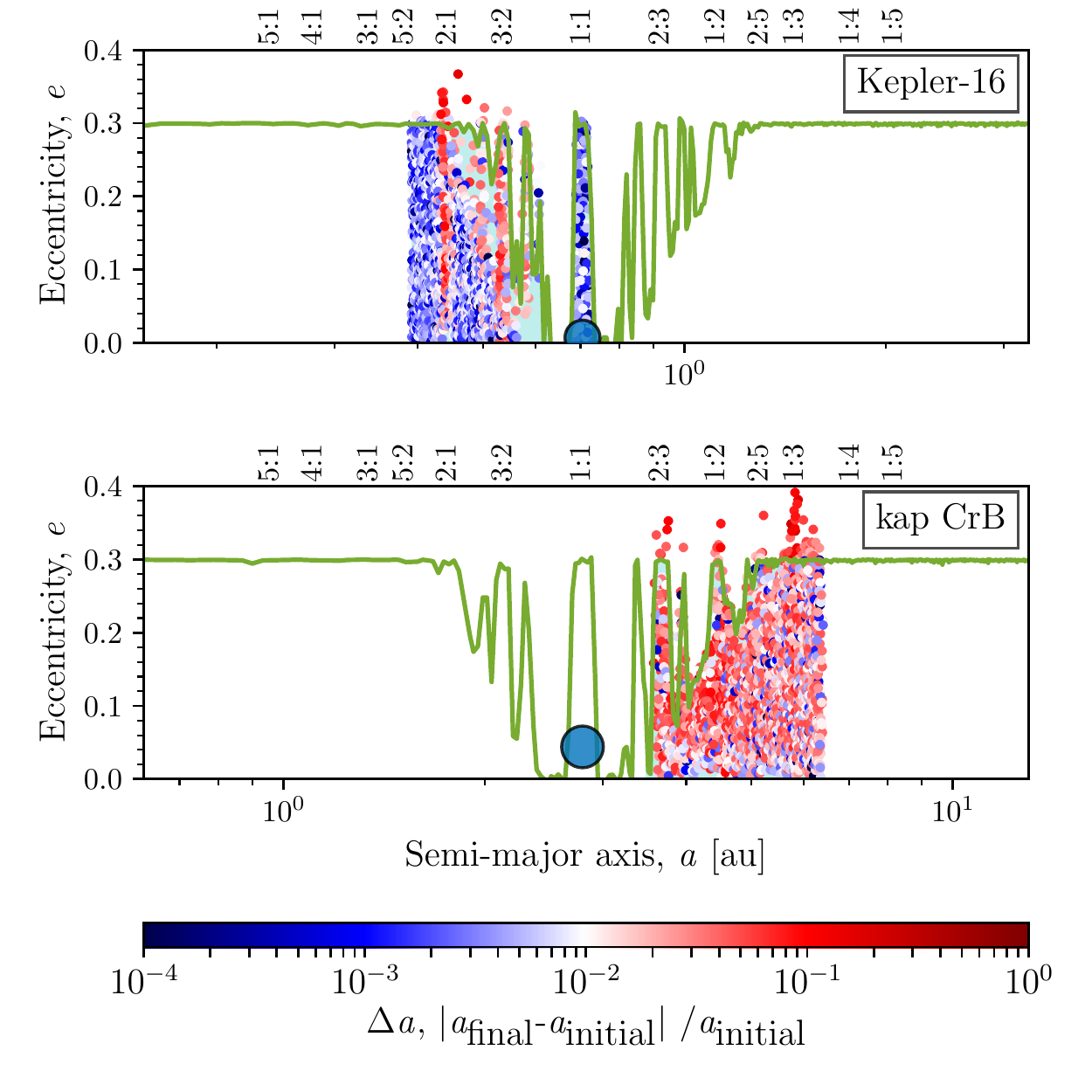} 
	\caption{A comparison between the predictions made using our method, and previously run numerical TP simulations conducted by \protect\cite{Agnew2018} for Kepler-16 ($e_\mathrm{pl}=0.0069$) and kappa Cr B ($e_\mathrm{pl}=0.044$). The coloured dots show the final position of the TPs in ($a$,$e$) space with the colour representing $\Delta a$. The blue circle represents the planet, the green curve shows its optimistic stability signature and the shaded green region shows the HZ below the stability signature. It can be seen that a large majority of the stable TPs fall below the stability signature, where we predict the unperturbed stable regions in the system to be.}\label{fig:tp_predictions}	
\end{figure}

\subsubsection{TP companion in HZ}\label{subsubsec:tp}
We first compare our predictions with systems that were simulated numerically with massless TPs. \cite{Agnew2018} conducted a large suite of simulations for 182 single Jovian planet systems using $5000$ TPs randomly distributed throughout the HZ, and simulated each system for $10^7$ yrs. Depending on the orbital parameters of the Jovian planet, these simulations took days of computational time. In contrast, our predictions can be performed in seconds. We compare the TPs that were not removed from the system by ejection or collision at the end of the simulation with the predicted unperturbed, stable regions below the stability signature using our method to assess the robustness of our approach for massless bodies.

As our stability signatures have only been determined for mass ratios $\mu \leq 1024 \textrm{M}_{\oplus}/\textrm{M}_\odot$ and $e\leq0.3$, this places limitations on which systems we can compare with. \cite{Agnew2018} present 10 near-circular systems (i.e. $e_\mathrm{pl}<0.05$) that have mass ratios which fall within this range, and for which they have explored the stable regions in the HZ using a swarm of massless TPs. These systems are ideal candidates to demonstrate how the stability signatures can predict regions of HZ stability. To do this, we: 1) normalise the system by scaling the mass of the star and the planet, 2) interpolate between the masses on the look-up map to obtain the optimistic stability signature of each planet, 3) translate the signature of each planet to its semi-major axis, and 4) overlay the stability signature over ($a$,$e$) alongside the TPs that survived to the end of the numerical simulations ran by \cite{Agnew2018}. We can then examine if the TPs align with the area beneath the stability signature which corresponds with the unexcited, stable regions in the system. A worked example of this process for Kepler-16 can be found in Appendix~\ref{sec:appendix_examples}. Figure~\ref{fig:tp_predictions} shows the stability signatures and surviving TPs as found by \cite{Agnew2018} for Kepler-16 and kappa~Cr~B, two systems where the known planet interacts significantly with the HZ. All 10 of the near-circular systems can be seen in Appendix~\ref{sec:all_hz_companions}.
 
In these figures, the coloured dots represent the final position and the
relative change in semi-major axis of the TPs that remained at the end of the simulation, the green curve is the stability signature, the shaded green area is the HZ that exists beneath the signature, and the blue point is the massive planet. It can be seen that there is strong agreement between the stability signatures and the surviving TPs for each system as was determined numerically by \cite{Agnew2018}. Especially so in the case of Kepler-16 as it is the most circular of the two systems. kappa~Cr~B also shows strong agreement in that a large majority of the TPs fall below the stability signature, but as the orbit of the planet is slightly less circular (0.0069 in Kepler-16 vs 0.044 in kappa~Cr~B) we see some excitation of the TPs to eccentric orbits greater than $e>0.3$. Regardless, for near-circular systems we see that the stability signature is a strong predictor of stable regions, and that further development to incorporate eccentric orbits has the potential to yield even stronger predictions for less circular orbits.

\begin{figure} 
    \centering
    \includegraphics[width=\linewidth]{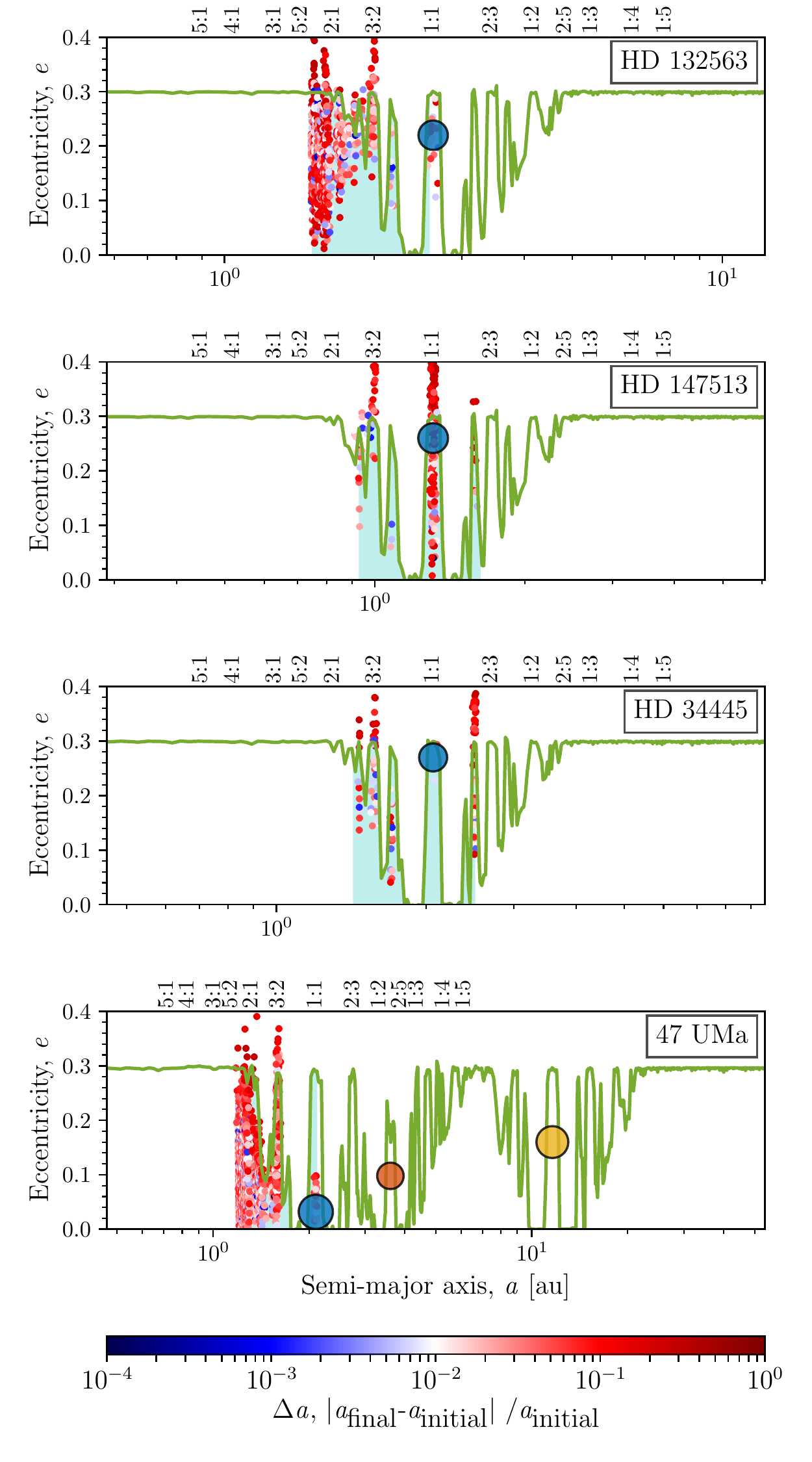}
	\caption{A comparison between the predictions made using our method, and previously run numerical $1~\textrm{M}_{\oplus}$ body simulations conducted by \protect\cite{Agnew2017} and \protect\cite{Agnew2018b} for HD~132563~B, HD~147513, HD~34445 and 47~UMa. The coloured dots show the final position of the $1~\textrm{M}_{\oplus}$ bodies in ($a$,$e$) space with the colour representing $\Delta a$. The large coloured circles represent the planets in each system, the green curve shows optimistic stability signature of each system and the shaded green region shows the HZ below the signature. It can be seen that a large majority of the stable $1~\textrm{M}_{\oplus}$ bodies fall below the optimistic stability signature, where we predict the unperturbed stable regions in the system to be.}\label{fig:massive_predictions}	
\end{figure}
\subsubsection{Earth-mass companion in HZ}
As one of the goals of our work is to identify where Earth-mass planets could maintain stable orbits, we need to demonstrate the stability predictions are not just valid for massless TPs, but also for massive bodies. We have the same mass ratio and eccentricity constraint as we did for the TP predictions, and so we similarly focus on systems with mass ratios $\mu \leq 1024 \textrm{M}_{\oplus}/\textrm{M}_\odot$ where the planet moves on near-circular orbits. \cite{Agnew2017} conducted a suite of simulations to explore the stable regions in the HZ of 15 single Jovian planet systems by sweeping a $1~\textrm{M}_{\oplus}$ body over the ($a$,$e$) parameter space. For each system, 20,400 individual simulations were run, where the orbital parameters of the $1~\textrm{M}_{\oplus}$ body were gradually varied to cover the desired parameter space. Such thorough numerical simulations take days to weeks to complete, whereas our predictions can be performed in seconds. We compare the $1~\textrm{M}_{\oplus}$ bodies that were not removed from the system by ejection or collision at the end of the simulations with the predicted unperturbed, stable regions below the stability signature using our method to assess the robustness of our approach for Earth-mass bodies.

None of those systems studied by \cite{Agnew2017} would be considered near-circular, so we just look at those that have mass ratios that fall within the $\mu$ range and with an $e_\mathrm{pl}\leq0.3$. Three systems fulfill these requirements, HD~132563~B, HD~147513 and HD~34445, and hence are candidates to demonstrate how the stability signatures can predict HZ stability. While these systems have relatively low eccentricities ($0.22$, $0.26$ and $0.27$ respectively), they are not what one would consider near-circular. As such we also include a multiple planet system where the planet in closest proximity to the HZ is much closer to circular. That system is 47~UMa, for which \cite{Agnew2018b} carried out a similar massive body parameter sweep. In this case the separations between planets in $47$~UMa are such that the predictions are still useful. To perform these HZ predictions we perform the same steps as outlined in section~\ref{subsubsec:tp}, but instead of plotting the TPs that survived to the end of the numerical simulations, we plot the $1~\textrm{M}_{\oplus}$ bodies that survived to the end of the numerical simulations performed by \cite{Agnew2017} and \cite{Agnew2018b}. Figure~\ref{fig:massive_predictions} shows the stability signatures and surviving $1~\textrm{M}_{\oplus}$ bodies found by \cite{Agnew2017} and \cite{Agnew2018b} for HD~132563~B, HD~147513, HD~34445 and 47~UMa. 

An additional step required for 47~UMa is that the stability signatures of all planets must be combined. At any semi-major axis location, the stability signature of each planet will have a corresponding eccentricity  representing the maximum eccentricity of the unexcited region. As we assess whether one planet would excite any other, the smallest of these eccentricity values will ultimately bound the unexcited region at that semi-major axis location. As such, we can take the lowest value of all the stability signatures at each semi-major axis location to generate a  combined stability signature. The combined signature for 47~UMa is that shown in the bottom panel of Figure~\ref{fig:massive_predictions}.

In these figures, the coloured dots represent the final position and the
relative change in semi-major axis of the $1~\textrm{M}_{\oplus}$ bodies that remained at the end of the simulation, the green curve is the stability signature, the shaded green area is the HZ that exists beneath the signature, and the blue, red and yellow points are the massive planets.

As the known massive planet in HD~132563~B, HD~147513 and HD~34445 is eccentric, the stable bodies do not fit the predictions as accurately as the near-circular cases. We can see that while the massive bodies tend to align along the semi-major axis, they are excited to higher eccentricity orbits. This matches what we see in the eccentric simulations as described in section~\ref{subsec:ecc} and with what can be seen in Appendix~\ref{sec:appendix_all_sims}. However, it does demonstrate that the predictive ability of the stability signatures is still valid for a $1~\textrm{M}_{\oplus}$ body. In the case of 47~UMa, as the planets are adequately separated the stability signature of the planet nearest to the HZ can be used to predict stability within it. That planet moves on a much more circular orbit, and so we can see that the stable $1~\textrm{M}_{\oplus}$ bodies align very closely with the area below the stability signature. This agreement of the stability signature with $1~\textrm{M}_{\oplus}$ bodies shows that our method can be used effectively to identify where stable, HZ Earth-mass planets could maintain stable orbits in known exoplanet systems without the need to run exhaustive numerical simulations.

It is important to note that Figures~\ref{fig:tp_predictions} and \ref{fig:massive_predictions} demonstrate that several bodies do remain stable at the end of the numerical simulations carried out by \cite{Agnew2017}, \cite{Agnew2018} and \cite{Agnew2018b} that have eccentricities greater than $0.3$. The limit of $e\leq0.3$ that we enforce here to find the stability signatures and develop our look-up map is the result of our focus on habitability \citep{Williams2002,Jones2005}. As such, it should be kept in mind that our method only predicts the unperturbed regions in ($a$,$e$) space with $e\leq0.3$, even though it is possible for orbits with higher eccentricities to be stable.

\section{Conclusions}
\label{sec:summary}
Numerical simulations are integral to assessing the detailed dynamics of planetary systems. However, alternative methods to classify system stability are beneficial in ensuring computational resources are efficiently allocated to assess only the most complicated systems. In this work, we have presented an alternative approach in assessing the stability of newly discovered exoplanet systems, such as those that will be found in coming years by \textit{TESS}, and its associated follow-up programs. This includes the dynamical stability of a multiple planet system with the best-fit orbital parameters, the overall stability of the HZ of a system, and whether an Earth-size planet could maintain a stable orbit within the HZ. The key findings of our work are as follows:
\begin{itemize}
	\item Mass ratio ($\mu=M_\mathrm{pl}/M_\mathrm{\star}$) and orbital eccentricity are very influential in determining a system's overall dynamical stability. In particular, we find that even moderate orbital eccentricities can prove to be destabilising, a finding that re-enforces the results of several studies that highlight the inverse relationship between multiplicity and eccentricity \citep{Carrera2016,Agnew2017,Zinzi2017}.
	\item ``Stability signatures'' can be obtained using our method for each planet in a multiple planet system, and these signatures can be used to assess the stability of each planet, and to determine the overall dynamical stability of a particular set of planetary parameters for a multiple planet system. Comparing our predictions with previously run numerical simulations using best-fit orbital parameters \citep{Agnew2018b}, we found our approach yields no strong disagreement in any of the systems assessed, and agreement in $32\%$ of cases.
	\item The stability of a planetary system can then be investigated by carrying out the stability signature assessment for many different permutations of planetary and orbital parameters. By considering the parameters over their respective error ranges, one can suggest whether more observations should be taken in order to better constrain the system to more stable orbital parameters.
	\item The stability signature interpolated from our results also proves effective at predicting the stability of massless TPs in near circular systems. Comparing the signature of several single planet systems with those TPs that were found to be stable with numerical simulations by \cite{Agnew2018} shows strong agreement.
	\item The stability signature is also good at predicting the stability of a $1~\textrm{M}_{\oplus}$ planet in the HZ of low eccentricity systems. Comparing the stability signature of several single planet systems with stable $1~\textrm{M}_{\oplus}$ bodies found numerically by \cite{Agnew2017}, we can see good agreement, with discrepancies being due to the stable $1~\textrm{M}_{\oplus}$ bodies being more excited to higher eccentricities as a result of the existing single planet not being on a near-circular orbit (an assumption of our model).
	\item The stability signature is also very good at predicting the stability of a $1~\textrm{M}_{\oplus}$ planet in multiple planet systems where the separation between known planets is such that the HZ only interacts with the nearest planet. $47$~UMa provides such a case where the planet nearest to the HZ is near circular, and here we again see strong agreement between the stability signature and the numerical simulations performed by \cite{Agnew2018b}.
\end{itemize}
Our work has focused on the simplest, near circular, co-planar case, and so there is room to refine our approach, which we intend to do in the future. We have demonstrated that our method shows a high degree of success for low eccentricity systems, and for multiple planet systems with large orbital separations.

This approach will be particularly useful for systems discovered with \textit{TESS}, allowing the system stability to be assessed for the best-fit orbital parameters and informing whether additional observations should be made to further constrain the orbits. It can also be used to rapidly predict which newly discovered systems may have dynamically stable Earth-size planets orbiting in their HZ. As our predictive tool does not require further numerical simulations, they can be incorporated into the Exoplanet Follow-up Observing Program for \textit{TESS} (ExoFOP-TESS) to provide these insights as planets are discovered.

\section*{Acknowledgements}
We wish to thank the anonymous referee for their thoughtful report, and helpful comments and recommendations that have improved the paper. We wish to thank Jessie Christiansen for helpful discussions regarding \textit{TESS} and ExoFOP-TESS. MTA was supported by an Australian Postgraduate Award (APA). This work was performed on the gSTAR national facility at Swinburne University of Technology. gSTAR is funded by Swinburne and the Australian Government's Education Investment Fund. This research has made use of the Exoplanet Orbit Database, the Exoplanet Data Explorer at exoplanets.org and the NASA Exoplanet Archive, which is operated by the California Institute of Technology, under contract with the National Aeronautics and Space Administration under the Exoplanet Exploration Program.

%%%%%%%%%%%%%%%%%%%%%%%%%%%%%%%%%%%%%%%%%%%%%%%%%%



\bibliographystyle{mnras}
\bibliography{project_4_final.bib} % if your bibtex file is called example.bib

%%%%%%%%%%%%%%%%%%%%%%%%%%%%%%%%%%%%%%%%%%%%%%%%%%

%%%%%%%%%%%%%%%%% APPENDICES %%%%%%%%%%%%%%%%%%%%%
\newpage\appendix
\section{Worked Examples}
\label{sec:appendix_examples}
\subsection{Multiple Planet Stability}
Here we show a worked example for assessing the dynamical stability of HD~5319. The current known arrangement of HD~5319 is summarised in Table~\ref{tab:hd_5319}. We follow the steps as they are outlined in section~\ref{app_stab}.

\begin{figure}
	\begin{subfigure}{\linewidth}
		\centering
    	\includegraphics[width=\linewidth]{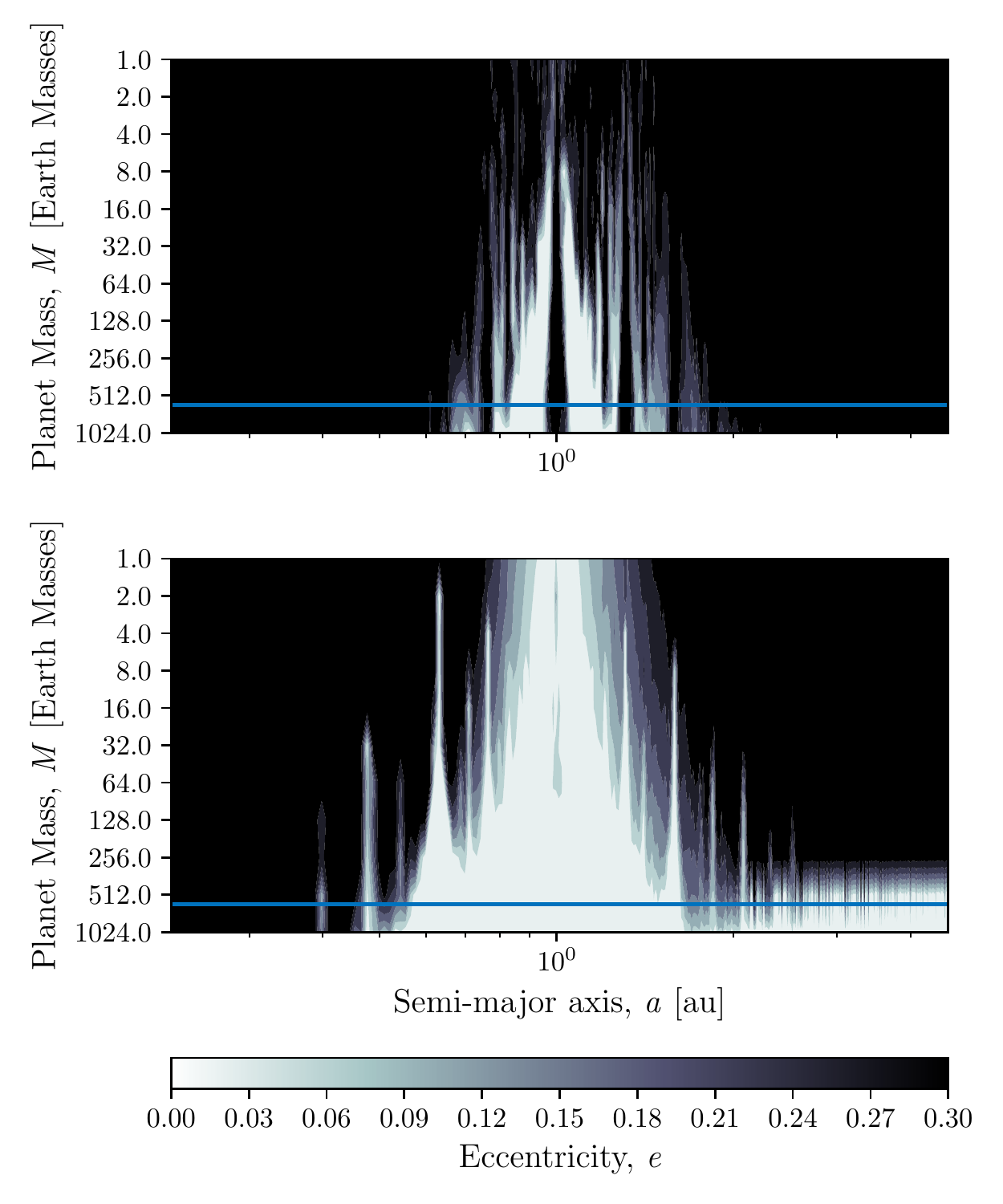}
    	\caption{Interpolate between simulated masses}\label{fig:stab_eg_cont}
	\end{subfigure}

	\begin{subfigure}{\linewidth}
		\centering
    	\includegraphics[width=\linewidth]{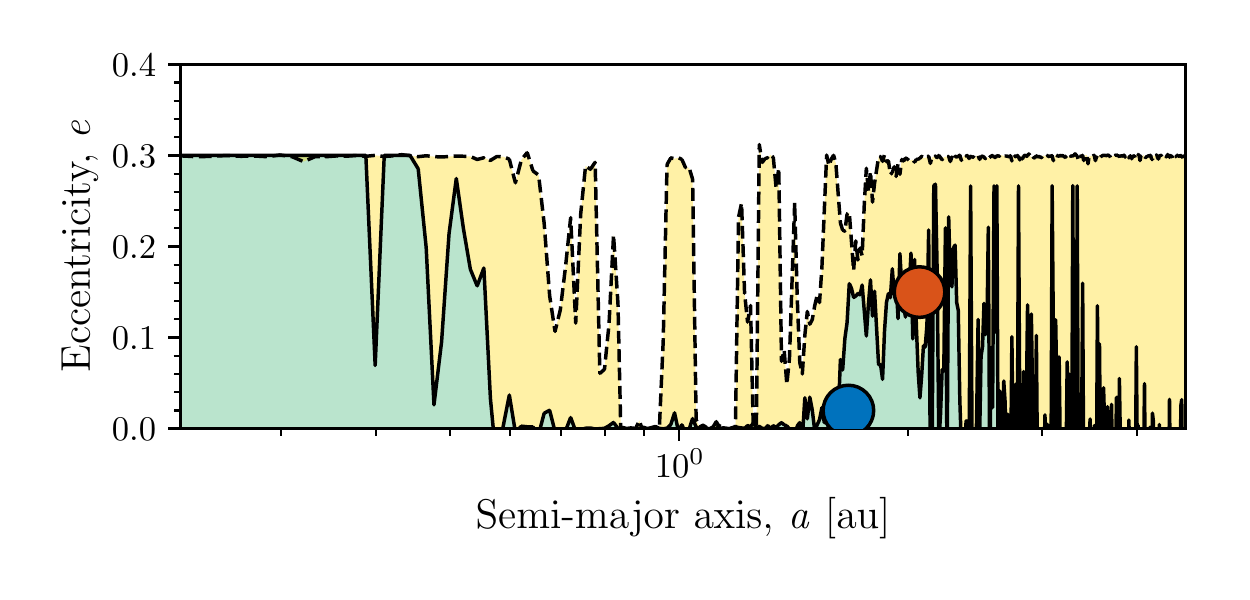}
    	\caption{Obtain stability signature from look-up maps}\label{fig:stab_eg_1}
	\end{subfigure}

%	\begin{subfigure}{\linewidth}
%		\centering
%    	\includegraphics[width=\linewidth]{images/examples/stab_eg_2.pdf}
%    	\caption{Translate stability signature}\label{fig:stab_eg_2}
%	\end{subfigure}

	\begin{subfigure}{\linewidth}
		\centering
    	\includegraphics[width=\linewidth]{images/predictions/stab_5319.pdf}
    	\caption{Translate stability signatures}\label{fig:stab_eg_3}
	\end{subfigure}

	%\begin{subfigure}{\linewidth}
	%	\centering
    %	\includegraphics[width=\linewidth]{images/examples/stab_eg_4.pdf}
    %	\caption{Assess multiple planety stability}\label{fig:stab_eg_4}
	%\end{subfigure}
	\caption{A worked example of HD~5319 demonstrating the key steps in assessing multiple planet stability. The coloured circles represent the planets, while the curves represent the stability signatures.}
\end{figure}

We first normalise the system. For a star with mass $M_\star=n \mathrm{M}_\odot$, we scale the masses of both the star, $M_\star$, and the combined mass of the planet pair,  $M_\mathrm{pl,1}'=_\mathrm{pl,2}'=M_\mathrm{pl,1}+M_\mathrm{pl,2}$, by $1/n$. This yields a normalised planet mass of $$\frac{1}{1.51} \ (1.76+1.15)~\mathrm{M}_\mathrm{Jup} = 1.927~\mathrm{M}_\mathrm{Jup} = 612.8~\mathrm{M}_\oplus$$ With the normalised mass, we are now able to obtain the stability signature of the each planet (which is the same in the two planet case) from our look-up maps by interpolating between the masses we have simulated. Figure~\ref{fig:stab_eg_cont} shows where the combined normalised mass of the planets ``slices'' across our two maps. The shading of the look-up map corresponds with the maximum unexcited eccentricity for the optimistic stability signature, and the minimum excited eccentricity for the conservative stability signature, each slice of the contour-map will have a corresponding curve. These are shown in Figure~\ref{fig:stab_eg_1}.

Having obtained the stability signatures for the planets, we next need to translate the signatures to the semi-major axes of each planet. The domain over which our simulations were run is $$a\in [0.215,4.642]$$ As per equation~\ref{eq:translate}, the domain of our translated signature will be $$a\in a_\mathrm{pl}\ [0.215,4.642]$$ This yields a domain for planet b of $$a\in 1.6697\cdot [0.215,4.642]$$ $$\implies a\in [0.359,7.751]$$ and a domain for planet c of $$a\in 2.071\cdot[0.215,4.642]$$ $$\implies a\in [0.445,9.614]$$ The translated signatures for each planet can be seen in Figure~\ref{fig:stab_eg_3}.

% To assess stability, we determine whether each planet excites the other by whether the planet falls above or below every other planet's stability signature. Instead of this, we can take the minimum value of the curve at each location and do one comparison to this combined stability curve. Figure~\ref{fig:stab_eg_3} shows the curves being combined.

To assess stability, we identify where each planet is located relative to the stability signatures of the other. Figure~\ref{fig:stab_eg_3} demonstrates that in the case of HD~5319, planet c falls outside the optimistic stability region of planet b, and so is excited and so it may be unstable.

\begin{table}
\caption{Summary of stellar mass, planetary mass and planetary semi-major axes of the planet/s in the HD~5319 and Kepler~16 systems as they are currently known.}
    \centering
    \begin{tabular}{c c c c}
        \toprule    
                & \textit{HD~5319}               & b & c \\
        %  \midrule
        %   & & & \\
         \midrule
         Mass   & $1.51~\mathrm{M}_\odot$      & $1.76~\mathrm{M}_\mathrm{Jup}$ & $1.15~\mathrm{M}_\mathrm{Jup}$ \\
         Normalised Mass & $1~\mathrm{M}_\odot$  & $1.166~\mathrm{M}_\mathrm{Jup}$& $0.762~\mathrm{M}_\mathrm{Jup}$ \\
         Semi-major axis [au] & -       & $1.6697$              & $2.071$ \\
         \midrule
                &\textit{Kepler-16} & b & \\
         \midrule
         Mass   & $0.69~\mathrm{M}_\odot$      &  $0.333~\mathrm{M}_\mathrm{Jup}$   & -    \\
         Normalised Mass & $1~\mathrm{M}_\odot$  &  $0.483~\mathrm{M}_\mathrm{Jup}$  & -    \\
         Semi-major axis [au] & -       & $0.7048$  & -      \\
         \midrule
    \end{tabular}
    \label{tab:hd_5319}
\end{table}

\subsection{Predicting HZ companions}
Here we show a worked example of predicting HZ companions in Kepler-16. The current known arrangement of Kepler-16 is summarised in Table~\ref{tab:hd_5319}. We follow the steps as they are outlined in section~\ref{app_stab}.

\begin{figure}
	\begin{subfigure}{\linewidth}
		\centering
    	\includegraphics[width=\linewidth]{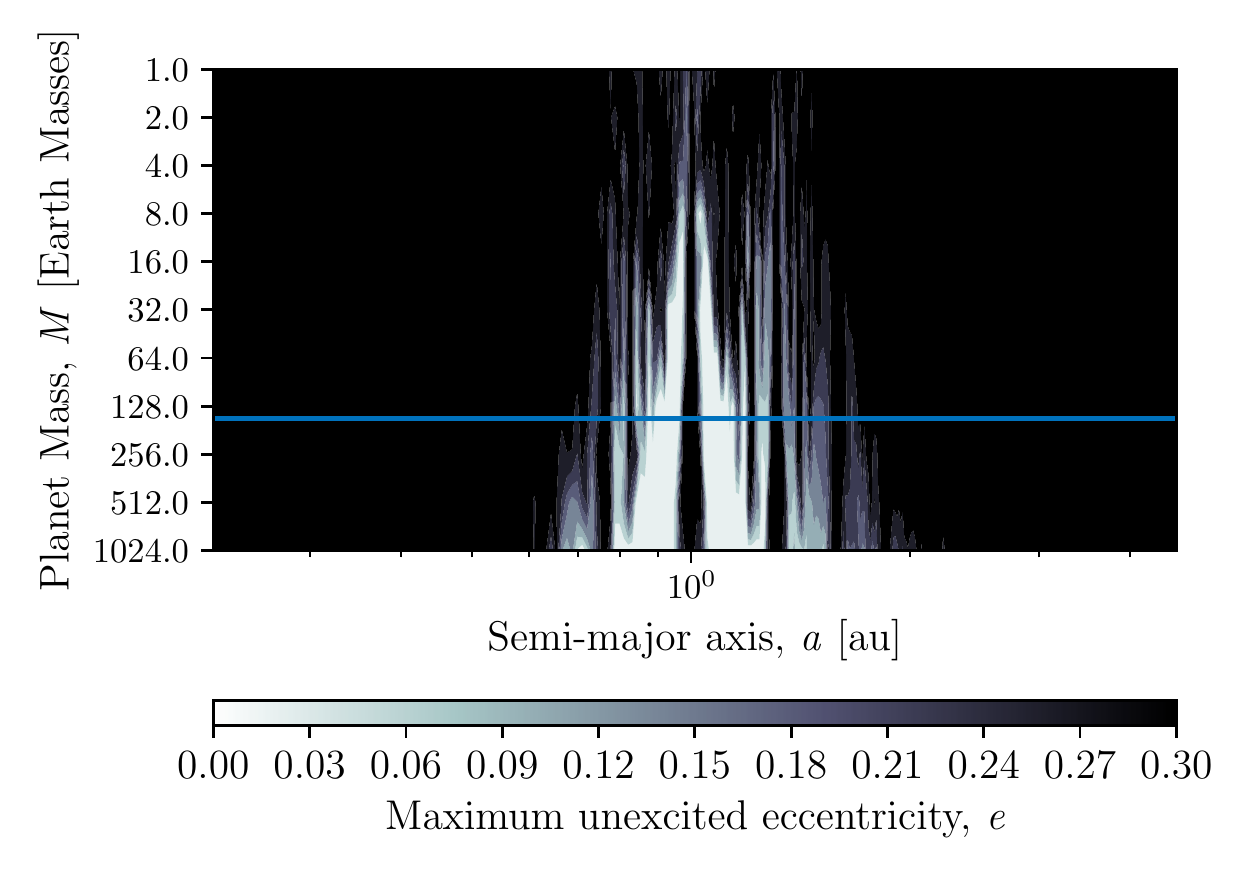}
    	\caption{Interpolate between simulated masses}\label{fig:hz_eg_cont}
	\end{subfigure}

	\begin{subfigure}{\linewidth}
		\centering
    	\includegraphics[width=\linewidth]{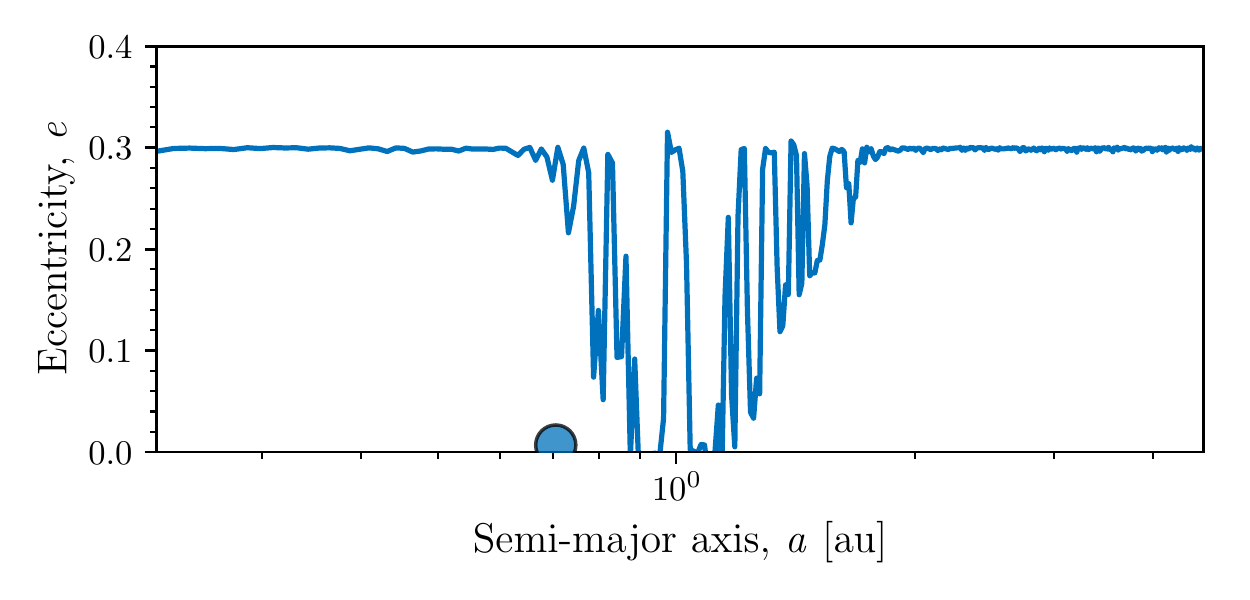}
    	\caption{Obtain stability signature from look-up map}\label{fig:hz_eg_1}
	\end{subfigure}

% 	\begin{subfigure}{\linewidth}
% 		\centering
%     	\includegraphics[width=\linewidth]{images/examples/tp_eg_2.pdf}
%     	\caption{Translate stability signature}\label{fig:hz_eg_2}
% 	\end{subfigure}

	\begin{subfigure}{\linewidth}
		\centering
    	\includegraphics[width=\linewidth]{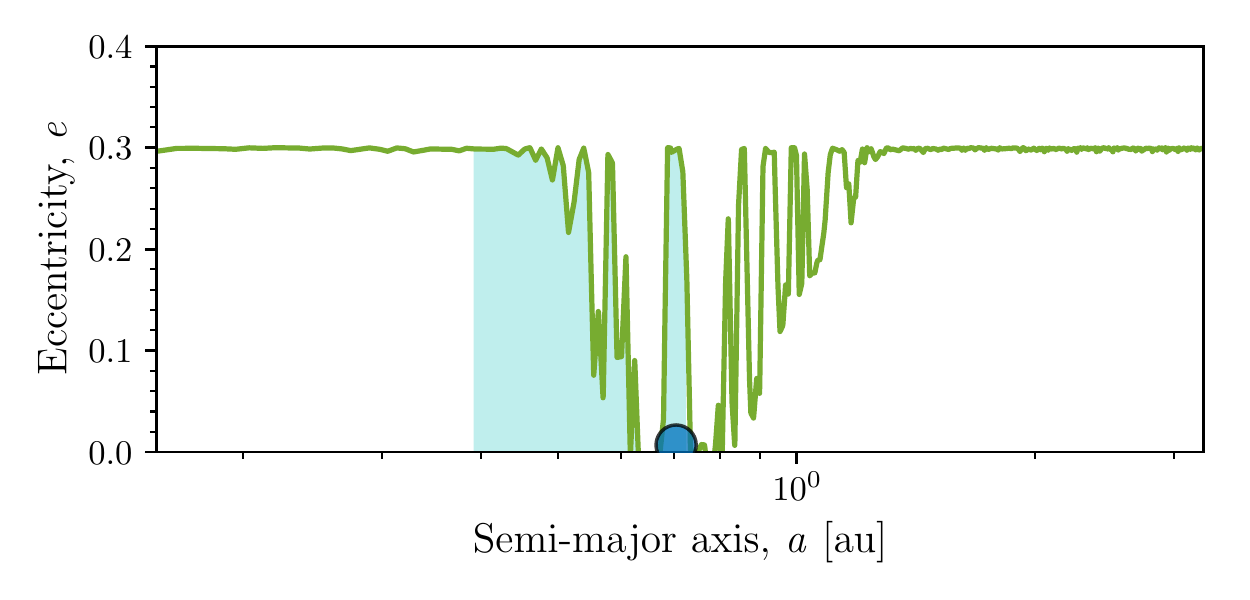}
    	\caption{Translate stability signature, compute HZ and overlay}\label{fig:hz_eg_3}
	\end{subfigure}

	\begin{subfigure}{\linewidth}
		\centering
    	\includegraphics[width=\linewidth]{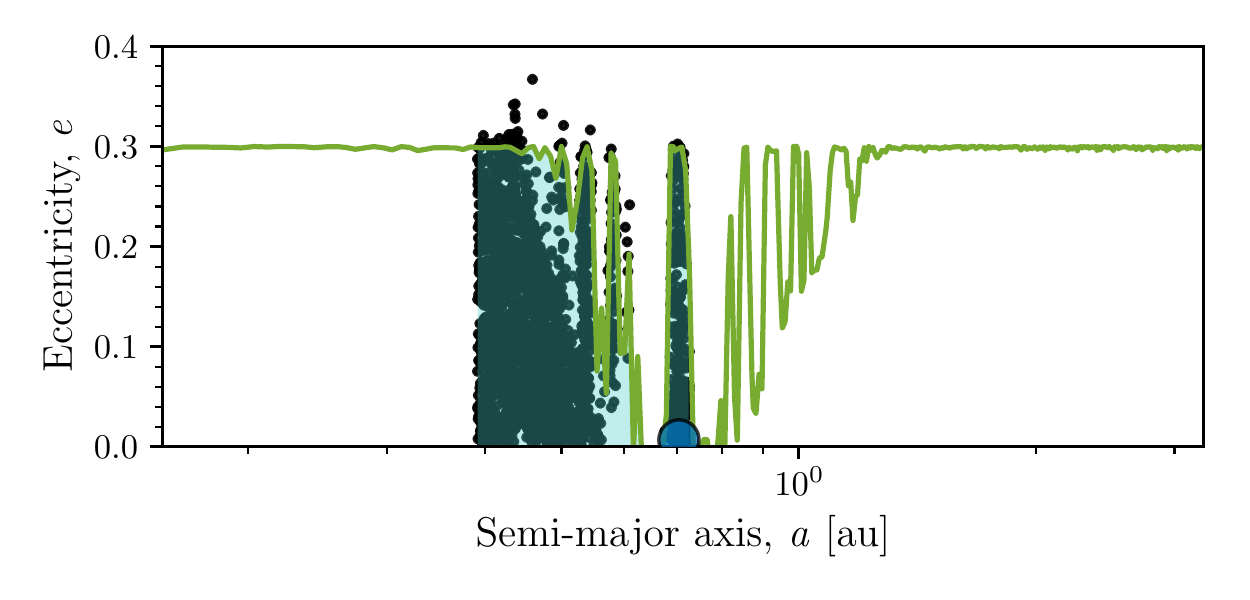}
    	\caption{Overlay previously run numerical results}\label{fig:hz_eg_4}
	\end{subfigure}
	\caption{A worked example of Kepler-16 demonstrating the key steps in determining if HZ companions may exist. The blue circle represents the planets, the black dots represent the massless TPs, and the curve represents the stability signatures.}
\end{figure}

We first normalise the system. For a star with mass $M_\star=n \mathrm{M}_\odot$, we scale the masses of both the star, $M_\star$, and planet, $M_\mathrm{pl}$, by $1/n$. This yields normalised mass for planet~b of $$\frac{1}{0.69} \ 0.333~\mathrm{M}_\mathrm{Jup} = 0.483~\mathrm{M}_\mathrm{Jup} = 153.4~\mathrm{M}_\oplus$$ With the normalised mass, we are now able to obtain the stability signature for planet~b from our look-up map by interpolating between the masses we have simulated. Figure~\ref{fig:hz_eg_cont} shows where the normalised mass of the planet ``slices'' across our map. For predicting HZ companions, we use the optimistic stability signature. The shading of the look-up map corresponds with the maximum unexcited eccentricity for the optimistic stability signature. This is shown in Figure~\ref{fig:hz_eg_1}.

Having obtained the stability signature of planet~b, we next need to translate the signature to its semi-major axis. The domain over which our simulations were run is $$a\in [0.215,4.642]$$ As per equation~\ref{eq:translate}, the domain of our translated signature will be $$a\in a_\mathrm{pl}\ [0.215,4.642]$$ This yields a domain for planet~b of $$a\in 0.7048 \cdot [0.215,4.642]$$ $$\implies a\in [0.152,3.272]$$ We then compute the HZ boundaries for the system. We use the conservative HZ definition for a $1~\mathrm{M}_\oplus$ as given in \cite{Kopparapu2014} for our HZ boundaries. Figure~\ref{fig:hz_eg_3} shows the translated signature and the HZ that falls below it for Kepler-16.

% The translated signature for planet~b can be seen in Figure~\ref{fig:hz_eg_2}.
Finally, using the results of the numerical simulations run by \cite{Agnew2018}, we can plot the final position of the TPs remaining at the end of their simulation in ($a$,$e$) space. Figure~\ref{fig:hz_eg_4} demonstrates there is tight agreement between the stability signatures and the numerical simulations.
% \begin{table}
% \caption{Kepler-16}
%     \centering
%     \begin{tabular}{c c c c}
%         \toprule    
%                 & Star                  & b  \\
%          \midrule
%          Mass   & $0.69~\mathrm{M}_\odot$      & $0.333~\mathrm{M}_\mathrm{Jup}$ \\
%          Normalised Mass & $1~\mathrm{M}_\odot$  & $0.483~\mathrm{M}_\mathrm{Jup}$\\
%          Semi-major axis [au] & -       & $0.7048$              \\
%          \midrule
%     \end{tabular}
%     \label{tab:kepler16}
% \end{table}
\section{HZ companion Figures}
\label{sec:all_hz_companions}
\begin{figure*}
    \centering
    \includegraphics[width=\linewidth]{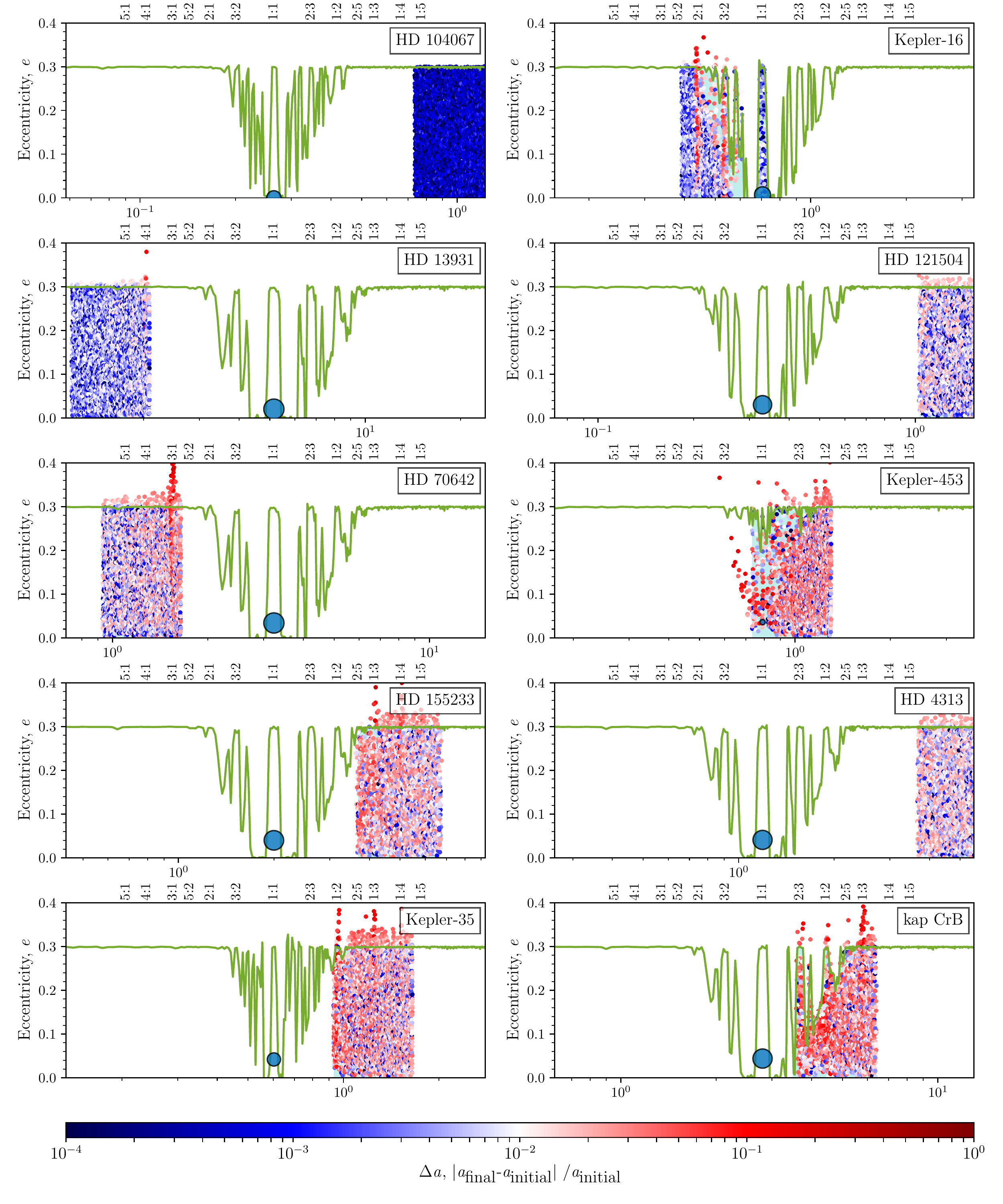}
    \caption{Comparison between all the HZ predictions made using our method, and previously run numerical TP simulations conducted by \protect\cite{Agnew2018} for all near-circular systems ($e_\mathrm{pl}<0.05$). The coloured dots show the final position of the TPs in ($a$,$e$) space. The blue circle represents the planet, the green curve shows its stability signature and the shaded green region shows the HZ below the stability signature. It can be seen that a large majority of the stable (coloured dots) TPs fall below the stability signature, where we predict the unperturbed stable regions in the system to be.}
\end{figure*}
\section{Test particle Figures}
\label{sec:appendix_all_sims}
\begin{figure*}
    \centering
    \includegraphics[width=\textwidth]{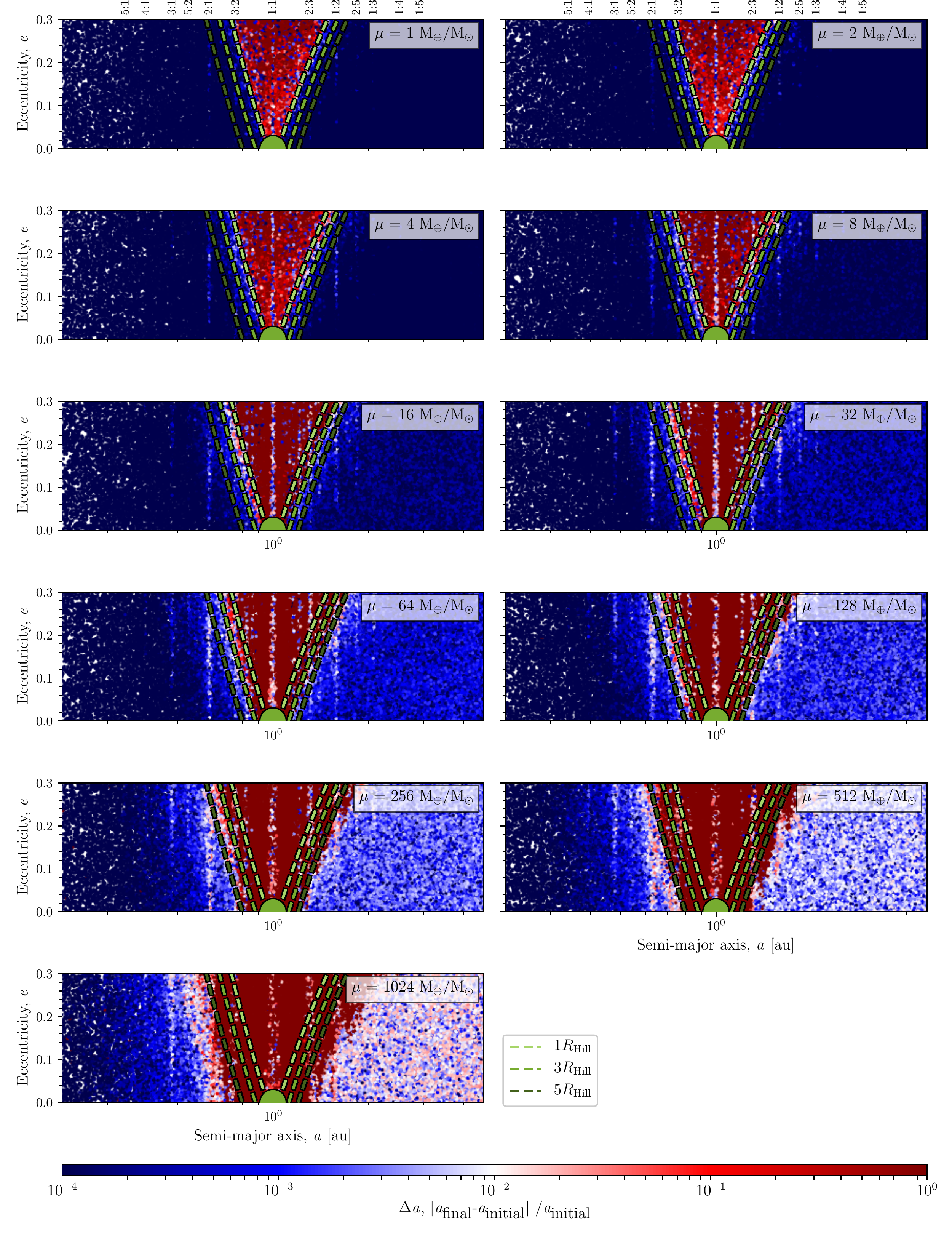}    
    \caption{The 11 simulations using different mass ratios to investigate its influence on stability near to the planet. The dots show the initial position of the $10^5$ TPs in ($a$,$e$) space, while the colour represents the relative change in semi-major axis. The shaded red curves show the boundaries of the 1, 3 and 5 Hill radii of the planet.}\label{fig:mass_effect}
\end{figure*}
\begin{figure*}
    \centering
    \includegraphics[width=\textwidth]{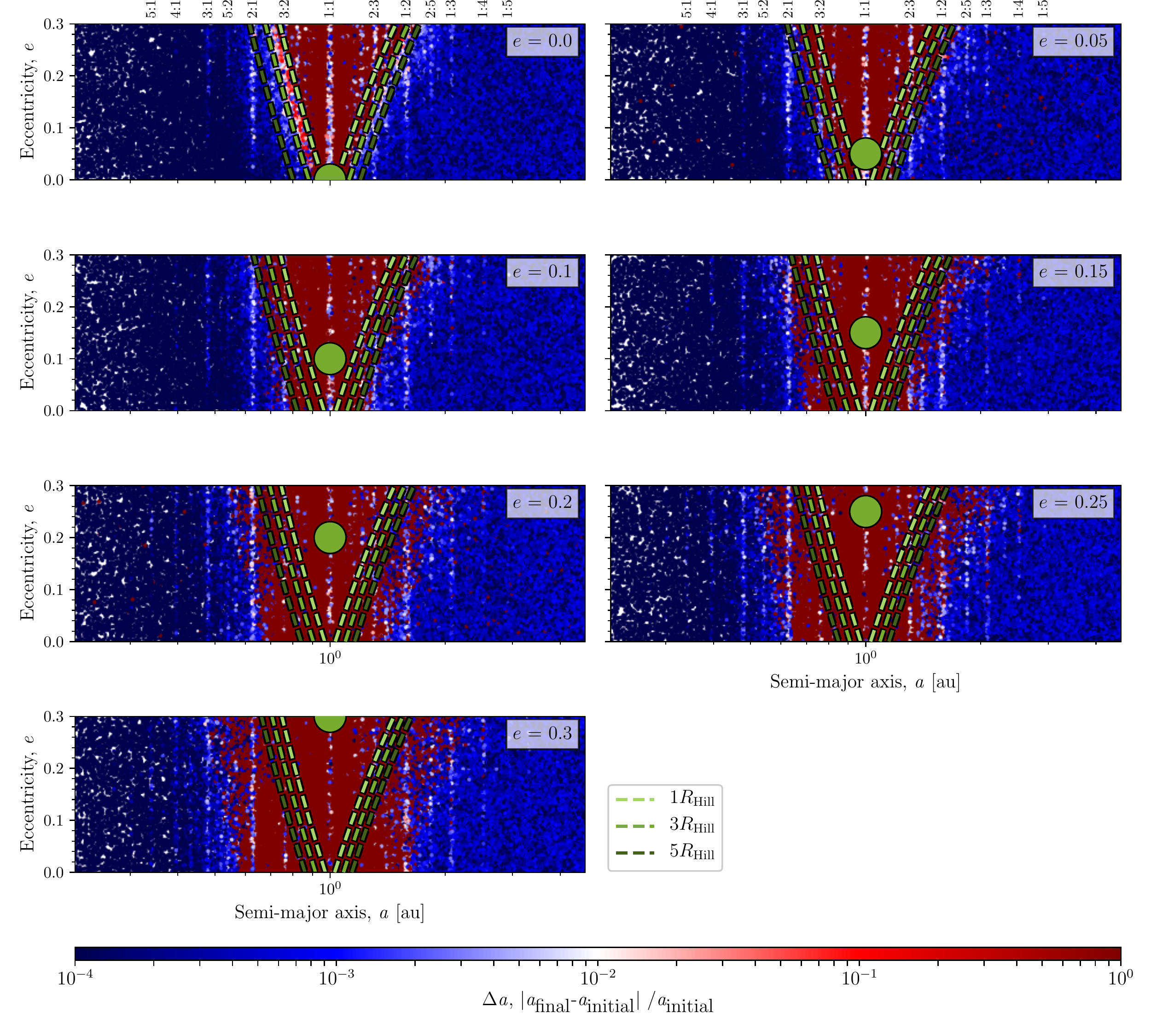}    
    \caption{The 7 simulations using different eccentricities to investigate its influence on stability near to the planet. The dots show the initial position of the $10^5$ TPs in ($a$,$e$) space, while the colour represents the relative change in semi-major axis. The shaded red curves show the boundaries of the 1, 3 and 5 Hill radii of the planet.}\label{fig:ecc_effect}
\end{figure*}
\begin{figure*}
    \centering
    \includegraphics[width=\textwidth]{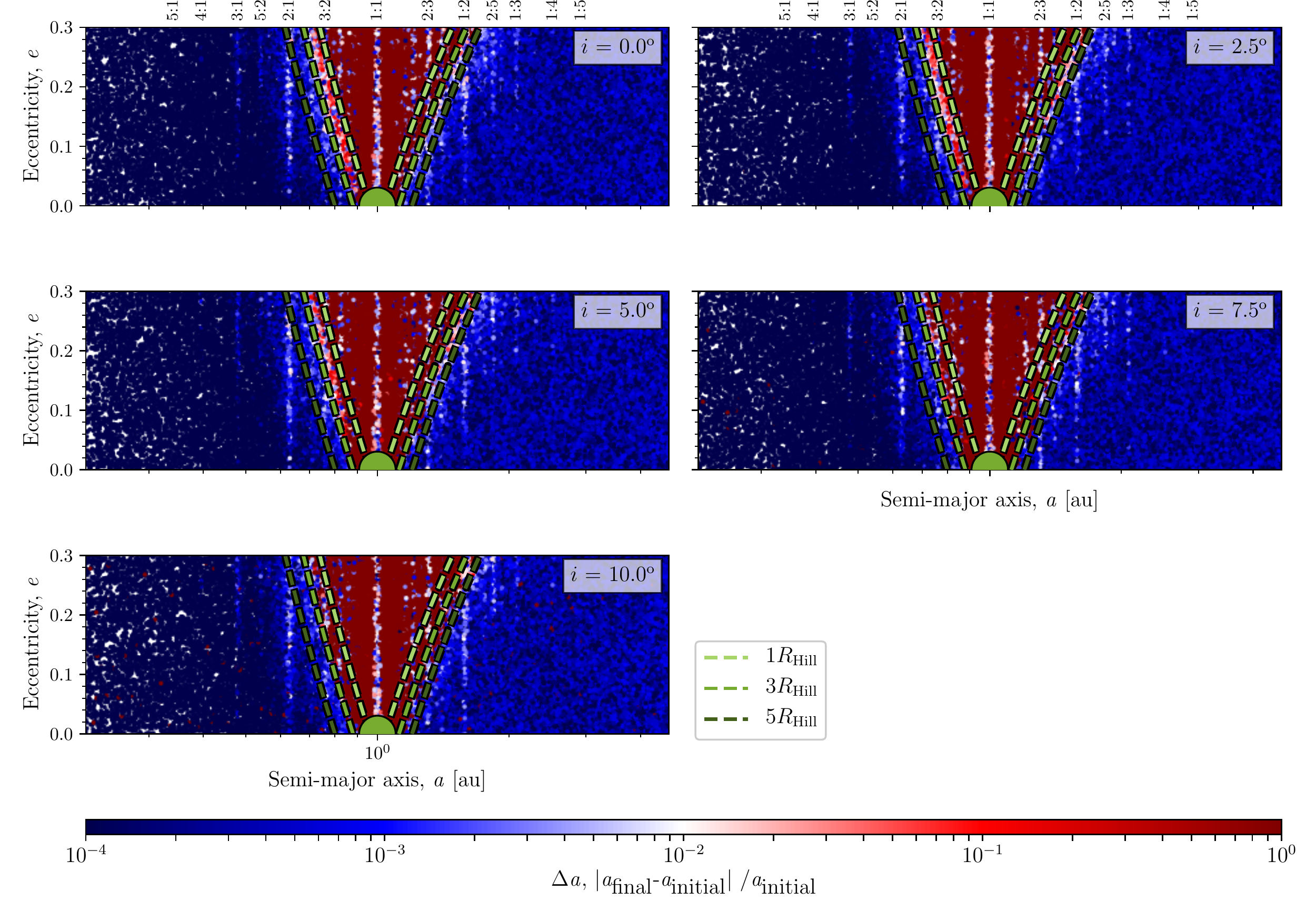}
    \caption{The 5 simulations using different inclinations to investigate its influence on stability near to the planet. The dots show the initial position of the $10^5$ TPs in ($a$,$e$) space, while the colour represents the relative change in semi-major axis. The shaded red curves show the boundaries of the 1, 3 and 5 Hill radii of the planet.}\label{fig:inc_effect}
\end{figure*}

%%%%%%%%%%%%%%%%%%%%%%%%%%%%%%%%%%%%%%%%%%%%%%%%%%

% Don't change these lines
\bsp	% typesetting comment
\label{lastpage}
\end{document}